\documentclass[a4paper,11pt]{article} \title{Off-Critical SLE(2) and
  SLE(4):\\ a Field Theory Approach.}
\author{}

 \newcommand{\vev}[1]{\langle
  #1 \rangle} \newcommand{\bvev}[1]{\big\langle #1 \big\rangle}

 \newcommand{\im} {\Im
  \textrm{m }} 
\newcommand{\re} {\Re \textrm{e }} 
 
\newcommand{\bH} {\mathbb{H}} 
\newcommand{\bC} {\mathbb{C}} 
\newcommand{\bR} {\mathbb{R}} 
\newcommand{\bD} {\mathbb{D}} 
\newcommand{\EX} {\mathsf{E}} 
\newcommand{\sL} {\mathcal{L}} 
\newcommand{\sO} {\mathcal{O}} 
\newcommand{\half} {\frac{1}{2}} 
 \newcommand{\ind} {\mathbf{1}} 
\newcommand{\ud} {\mathrm{d}} 
\newcommand{\corrl} {\zeta} 
\newcommand{\RW} {\mathrm{RW}} 
\newcommand{\Bplan} {\bf{B}} 
\newcommand{\BM} {\mathrm{BM}} \def\debut{\begin{eqnarray}}
  \def\fin{\end{eqnarray}} \def\non{\nonumber}

\usepackage[final]{graphicx}
\usepackage{amssymb} 
\usepackage{amsmath}
\usepackage{oldgerm} 
\usepackage{verbatim} \input{epsf}

\begin{document}
\maketitle

\vspace{-0.5 truecm}

\centerline{\large Michel Bauer\footnote[1]{ Institut de Physique
    Th\'eorique de Saclay, CEA-Saclay, 91191 Gif-sur-Yvette, France
    and Laboratoire de Physique Th\'eorique, Ecole Normale
    Sup\'erieure, 24 rue Lhomond, 75005 Paris, France.  {\small \tt
      <michel.bauer@cea.fr>}}, Denis Bernard\footnote[2]{Member of the
    CNRS; Laboratoire de Physique Th\'eorique, Ecole Normale
    Sup\'erieure, 24 rue Lhomond, 75005 Paris, France.  {\small \tt
      <denis.bernard@ens.fr>}} and Luigi
  Cantini\footnote[3]{Laboratoire de Physique Th\'eorique, Ecole
    Normale Sup\'erieure, 24 rue Lhomond, 75005 Paris, France.
    {\small \tt <luigi.cantini@ens.fr>}} }

\vspace{.3cm}

\vspace{1.0 cm}

\begin{abstract}
  Using their relationship with the free boson and the free symplectic
  fermion, we study the off-critical perturbation of SLE(4) and SLE(2)
  obtained by adding a mass term to the action. We compute the
  off-critical statistics of the source in the Loewner equation
  describing the two dimensional interfaces.  In these two cases we
  show that ratios of massive by massless partition functions,
  expressible as ratios of regularized determinants of massive and
  massless Laplacians, are (local) martingales for the massless
  interfaces.  The off-critical drifts in the stochastic source of the
  Loewner equation are proportional to the logarithmic derivative of
  these ratios. We also show that massive correlation functions are
  (local) martingales for the massive interfaces. In the case of
  massive SLE(4), we use this property to prove a factorization of the
  free boson measure.

\end{abstract}

\newpage


\section{Introduction}

Schramm-Loewner evolution (SLE) has been introduced to deal with
conformally invariant random curves. These curves may for instance be
thought of as interfaces in two dimensional critical statistical
systems. SLE is by now a (if not ``the'') standard tool to efficiently
formulate questions concerning these curves and, in many simple but
important cases, to get the answer by routine computations.  See
\cite{SLElawler, BBreport, CardySLE} for detailed introductions to
SLE.

Having reached this depth of insight in the critical case, it is a
natural question to wonder how SLE measures are deformed when
considering interfaces in statistical models not at the critical point
but slightly away from it (still in the scaling regime). Interfaces
out of criticality are at the moment very poorly understood to say the
least. There are several reasons to invest some efforts in this, some
of them more mathematical and some more physical, though the frontier
is fuzzy.

The first obvious remark is that interfaces, or domain walls, are
macroscopic structures that can be (and are) observed more directly
than microscopic correlations (though the average over the sample of a
local magnetization is the first accessible observable).  So they are
interesting to study for their own sake. A second obvious remark is
that some interface models are purely geometric. The canonical example
is percolation, for which all non-trivial observables are non local
and deal with cluster correlations.  But even if there are local
observables described by a local quantum field theory, this alone does
not yield a straightforward description of interfaces: local
quantum field theory does not deal easily with extended objects. As an
example, the fields in the Kac table for minimal conformal field
theories form a closed algebra but other fields which are non-local
with respect to this algebra are crucial to describe the interface.
SLE and its perturbations provide a framework to decipher properties
of such non-local excitations. A situation where this should be
relevant is the $O(n)$ model: one can see the introduction of defects
on the boundary to force the existence in the statistical mechanics
system of macroscopic interfaces pinned at special points as a trick
to get rid of some of the difficulties involved in the direct study of
a gas of loops at all scales in the continuum.

Of interest both to the physicist and the mathematician is the concept
of change of measure. In quantum field theory, the starting point is
often a formal measure $\ud \mu_S\equiv {\mathcal D} \varphi \,
e^{-S(\varphi)}$ whose rigorous construction is usually a formidable
challenge. The action $S(\varphi)$ may depend on parameters and
changing these parameters leads to families of measures.  For
instance, if $S=S_0+\int \lambda \mathcal{O} (\varphi)$ is a
perturbation of $S_0$ by some operator $\mathcal{O} (\varphi)$ one
gets formally that $\ud\mu_S=\ud\mu_{S_0}e^{-\int \lambda \mathcal{O}
  (\varphi)}$, or in a more measure-theoretic language, that $e^{-\int
  \lambda \mathcal{O} (\varphi)}$ is the Radon-Nikodym derivative of
$\mu_S$ with respect to $\mu_{S_0}$.  In general quantum field theory,
it is well-know that there is something poisonous in this statement
due to renormalization but a rigorous analysis is mostly out of reach.
However the framework of interfaces growth is a playground were
analogous questions can be tackled. By this we mean two things: first
that for some concrete models (see below) one can prove that some
interfaces measures have or do not have Radon-Nikodym derivatives with
respect to others, and second that when the answer is no for two
interfaces measures, it is also no for the two measures on local
degrees of freedoms that induce the interface measures.  To elaborate
on this issue, we need to introduce some background.

Consider first a finite system described by statistical mechanics.
Each configuration has a Boltzmann weight, usually strictly positive,
which may depend on some continuous parameters. Then two measures
corresponding to different values of the parameters have a
well-defined Radon-Nikodym derivative. In the thermodynamic (a
fortiori in the continuum) limit, this is another matter. In general,
a positive measure $\ud \mu$ is said to be absolutely continuous with
respect to another positive measure $\ud \mu_0$ on the same space if
for any set $B_0$ such that $\mu_0(B_0)=0$ there is a set $B \supset
B_0$ such that $\mu(B)=0$, more loosely, if negligible sets for
$\mu_0$ are also negligible for $\mu$. Under a technical condition,
this ensures that there is a $\mu_0$-measurable function $f$, called
the Radon-Nikodym derivative of $\mu$ with respect to $\mu_0$, such
that $\ud \mu=f\, \ud \mu_0$.  The theorem is obvious for finite or
countable spaces but delicate in general, see e.g.
\cite{Dudley,Rudin}. Observe that one does not look at individual
configurations (which have zero measure usually) but at subsets of
configurations. So the issue is that when going to the thermodynamic
limit, a subset of configurations can carry a finite weight for a
first choice of parameters but a vanishing weight for a second one.
Then the first measure is said to be singular with respect to the
second.

Suppose now that the system has boundary conditions imposing the
presence of one or several interfaces. From the point of view of
statistical mechanics, one could obtain the measure on interfaces as
the outcome of summing all the Boltzmann weights of configurations of
local degrees of freedom leading to a given position of the interface,
and the same holds in the continuum limit if we assume that interfaces
still make sense in that limit. If the measure on local degrees of
freedom depends on parameters, so does usually the measure on
interfaces. Assume that $\mu$ and $\mu_0$ are measures on local
degrees of freedom for two sets of parameters, and $\nu$ and $\nu_0$
are the corresponding measures on interfaces. Then if a set $I$ of
interfaces has measure $0$ for $\nu$, the set $B$ of local
configurations leading to an interface in $I$ has measure $0$ for
$\mu$. So if $\mu$ has a Radon-Nikodym derivative with respect to
$\mu_0$, so does $\nu$ with respect to $\nu_0$. The opposite does not
need to hold.

Along these lines a spectacular result has been obtained in
ref.\cite{NW}. For site percolation on the 2d triangular lattice
(where each site is occupied with probability $p$ and empty with
probability $1-p$ independently of the other sites) the measure
describing interfaces in the off-critical continuum limit (if $a$ is
the lattice mesh, one lets $p$ go to the critical value $p_c=1/2$
while keeping $(p-p_c)a^{-3/4}$ fixed) is singular with respect to the
critical interface measure. By the above remark, this also entails
that the local measures on hexagons are singular in the continuum
limit, a fact that can be understood as follows. The typical
fluctuation of the number of occupied sites is $\sim a^{-1}$ because
there are $\sim a^{-2}$ independent sites. However if $(p-p_c)\sim
a^{3/4}$, the typical asymmetry is $\sim a^{3/4}a^{-2}=a^{-5/4}$,
which is much larger that the fluctuation, so one can assert with
certainty that an individual sample is critical or not. In fact, the
same counting implies that on any set containing $\sim a^{-d}$
hexagons, the asymmetry $\sim a^{3/4} a^{-d}$ is much larger than the
fluctuation $\sim a^{-d/2}$ if $d > 3/2$ . The critical percolation
interface is bounded by $\sim a^{-7/4}$ hexagons so it covers enough
of the sample to feel a macroscopic effect of the tiny bias out of
criticality. Of course, this is cheating because the interface as a
set is correlated to the hexagon configuration. But this leads to
expect that along a typical interface sample out of criticality the
asymmetry between occupied and empty sites causes a systematic excess
of turns in one direction with respect to the other which is larger
than what could be attributed to fluctuations. This is the intuitive
basis for the result in \cite{NW}, but the actual proof involves
subtleties that are well beyond the scope of the above intuition.

The theorem in \cite{NW} is proved essentially without any recourse to
SLE and stochastic processes. However, SLE and more generally
stochastic processes can also provide relevant tools to address these
matters. Computing the interface measure from the local Boltzmann
weights is a very hard task, even at a critical point, and the
approach of SLE was to study measures on interfaces, defined directly
in the continuum, under two conditions, conformal invariance
(covariance under conformal transport to go from one domain to
another) and the domain Markov property (which asserts that the
probability distribution of the curves in a domain conditioned on an
initial portion of the curves is identical to the probability
distribution of the curves but in the domain minus the portion on
which we condition). This analysis led Schramm to the classification
-- in a one parameter family usually indexed by $\kappa \in
[0,+\infty[$ -- of conformally invariant measures on random curves
drawn on simply connected planar domains.  This should be contrasted
with the present status of conformal field theory, were several
mathematical axiomatics have been proposed but no general
classification is in view.

One of the remarkable features of Schramm's approach and result is
that the measure on critical curves can be realized in a natural way
as a 1d Brownian motion measure.  This goes via a construction of
interfaces via a stochastic growth process : to any continuous non
self-crossing curve joining two boundary points of a domain of the
complex plane one can associate, via a trick discovered by Loewner
involving a refinement of the Riemann mapping theorem, a real
continuous function $\xi_t$ for $t \in [0,+\infty[$. When the curves
in the domain have the statistics of a critical interface model, there
is a $\kappa \in [0,+\infty[$ such that $\xi_t/\sqrt{\kappa}$ is a
Brownian motion $B_t$. In particular the $\xi_t$ corresponding to a
critical interface measure is a Markov process.

Suppose now that the curves in the domain are non-critical
interfaces\footnote{That is, interfaces in a system out of criticality
  but in the scaling region. See Sec.  \ref{sec:dmlerw} for an example
  illustrating the passage to the continuum limit in the case of
  loop-erased random walks.}.  For such measures, conformal invariance
is broken i.e. the measures are not transported trivially by conformal
transformations in a change of domain. This is because going out of
criticality introduces a scale in the system, the correlation length
$\corrl$. But the domain Markov property, which has its roots in the
locality of the underlying statistical mechanics system, usually
survives.  The Loewner trick can still be used to associate to each
realization of the interface a continuous function $\xi_t$ for $t \in
[0,+\infty[$, and one of the ways to have a description of the
interface measure would be to give the measure on $\xi_t$.

There is little doubt that the probability that the interface has a
certain topology with respect to a finite number of points in the
domain should depend smoothly on the correlation length $\corrl$.
There is a small subtlety here: a collection of consistent finite
dimensional distributions fixes the law of a process, but the fact
that all finite dimensional distributions depend smoothly on $\corrl$
does not imply that the measures of the corresponding processes depend
smoothly on $\corrl$. Some observables, like the fractal dimension,
are computed by using an infinite number of points but are
nevertheless expected to be local enough to remain the same out of
criticality, in particular they depend smoothly on $\corrl$. On the
other hand, the construction of \cite{NW} introduces an observable
that depends on infinitely many points on the interface (but on
arbitrarily small segments) and uses it to prove that the measures at
criticality and out of criticality are not mutually absolutely
continuous.

At scales much smaller that the correlation length, i.e. in the
ultraviolet regime, the deviation from criticality is small and, as a
function of $t$, the off-critical $\xi_t$ is expected to share some
local features with its critical counterpart. This raises the question
whether $\xi_t$ can be decomposed as $\xi_t=\sqrt{\kappa}B_t + A_t$,
i.e. as the sum of a Brownian motion (scaled by $\sqrt{\kappa}$) plus
some process $A_t$ whose precise regularity would remain to be
understood but at least tamer than $B_t$ on small scales.

The results from \cite{NW} imply that there must be a problem with
that decomposition for percolation. The authors in \cite{NW} argue
that the decomposition $\xi_t=\sqrt{6}B_t + A_t$ should exist ($\kappa
=6$ for percolation), but that $A_t$ is too wild to expect absolute
continuity. Stochastic calculus for continuous stochastic processes
deals with processes $X_t$ that are called in the probabilistic jargon
semi-martingales. This means that they are defined as functionals of a
Brownian motion $B_t$ and have the following properties. First, $X_t$
is causal (mathematicians say adapted) in the sense that $X_t$ can be
computed knowing only $\{B_s, s \in [0,t]\}$ roughly speaking.
Second, there is a splitting $X_t=M_t + A_t$ as a sum where $M_t$ is a
stochastic integral\footnote{The stochastic integral $\int_0^t Y_s \ud
B_s$ is well-defined if the process $Y_t$ is causal and $\int_0^t
Y_s^2 ds$ is almost surely finite. Then the stochastic integral itself
is also causal. Note that $\int_0^t Y_s \ud B_s$ does not need to be a
martingale (i.e.  to be conserved in average) because it can get too
large. However it becomes a martingale if it is stopped as soon as its
absolute value reaches $n$ for $n=1,2,\cdots$.  So it is called a
\textit{local} martingale, a term we sometimes use in the sequel. Thus
being bounded is a sufficient condition for the process $\int_0^t Y_s
\ud B_s$ to be a martingale.  A milder useful criterion for being a
true martingale is that $\EX{\int_0^t Y_s^2 ds} < +\infty$.} $\int_0^t
Y_s \ud B_s$ and $A_t$ is of locally finite variation, i.e.  $\sup
\sum_i |A_{t_{i+1}}-A_{t_i}| < \infty$ when the $\sup$ is taken over
all subdivisions $0=t_0 < t_1 \cdots t_n=t$ for fixed $t$. The
separation of scales (very crudely, the variation of $M_t$ is of order
$\sqrt{dt}$ while that of $A_t$ is of order $dt$) between the two
terms implies that such a decomposition, if it exists, is unique. In
our case, we would have $M_t=\sqrt{\kappa}B_t$. If $A_t$ is regular
enough, the measures on the processes $\xi_s$ and $\sqrt{\kappa}B_s$,
$s \in [0,t]$ are absolutely continuous with respect to each other.
But for percolation $A_t$ is conjectured to be too wild.

Now look at scales large compared to $\corrl$. In this 
regime, the behavior is different and the interface should look like
another SLE with a new $\kappa _{ir}$. Take the Ising model as an
example. At criticality $\kappa=3$ but if the temperature is raised
above the critical point, general renormalization group arguments
indicate that at large scale the interface looks like the interface at
infinite temperature. From the explicit example of the hexagonal
lattice (plus maybe some confidence in universality) this limit is
percolation and $\kappa _{ir}=6$. But the infrared regime is
never attained in a bounded domain. 

Let us close this long introduction by stressing again that conformal
covariance and the domain Markov property have a rather different
status. Whereas conformal invariance emerges (at best) in the
continuum limit at criticality, the domain Markov property makes sense
and is satisfied on the lattice without tuning parameters for many
systems of interest. It can be considered as a manifestation of
locality (in the physicists terminology). Hence the domain Markov
property is still expected to hold off criticality.  However the
consequences of this property on $\xi_t$ do not seem to have a simple
formulation. As for conformal ``covariance'', there is a trick to
preserve it formally out of the critical point : instead of perturbing
with a scaling field $O(z,\bar{z})$ times a coupling \textit{constant}
$\lambda$, one perturbs by a scaling field times a density
$\lambda(z,\bar{z})$ of appropriate weight, in such a way that
$\lambda (z,\bar{z}) O(z,\bar{z}) \ud \bar{z}\wedge \ud z$ is a
$2$-form. If $\lambda(z,\bar{z})$ has compact support, one also gets
rid of infrared divergences that occur in unbounded domain. We shall
use this trick in some places, but beware that if perturbation theory
contains divergences, problems with scale invariance will arise, hence
the cautious word "formally" used above.

\section{Summary}

We are now in position to give a summary of our approach and results.

Besides ref. \cite{NW}, a few works on off-critical SLE have already
appeared but the study of this problem is still in his infancy. In
ref.\cite{BBK}, we exposed a possible framework for dealing with
deformations of SLE adapted to off-critical perturbations of the
underlying statistical models. This approach links off-critical SLE to
off-critical partition functions and field theories. It was
perturbatively applied, to first order in the perturbing mass only, to
off-critical loop erased random walks (LERW).  The aim of this paper
is to develop this method for two simple off-critical SLE, namely
massive SLE(2) and massive SLE(4).  These perturbations are simple
enough to be treated non perturbatively. Apparently some unpublished
related work on similar perturbations of SLE has been reported in
ref.\cite{MS}. There is no doubt that the perturbation of the Ising
model by the energy operator, corresponding to a shift of the
temperature, is amenable to the same techniques.  These three cases
(corresponding to certain perturbations of $\kappa=2,3,4$ i.e. central
charge $c=-2,1/2,1$ all correspond to free field theory and this is
the crucial point for our approach because it leads to computations of
(variations of) determinants.

For more general cases, the situation is less favorable.  Basic rules
of CFT fix unambiguously the process $A_t$ alluded to before to first
order in perturbation theory but not to second order and beyond.
Indeed, for the computation $\bvev{e^{-\int \lambda (z,\bar{z})
    O(z,\bar{z}) \ud \bar{z}\wedge \ud z}}_{bc}$ at order $n$ in
$\lambda$, one needs first to evaluate $\bvev{ O(z_1,\bar{z}_1) \cdots
  O(z_n,\bar{z}_n)}_{bc}$, which involves two boundary changing
operators but $n$ bulk fields. So if $n= 1$ the differential equation
coming from the fact that the boundary fields are degenerate at level
$2$ is enough to fix (almost) everything. But if $n >2$ a detailed
knowledge of the operator algebra of the theory, i.e. which states are
allowed as intermediate states in a correlator, is required. One could
restrict to perturbations of minimal models by minimal operators. Then
the value of $\bvev{ O(z_1,\bar{z}_1) \cdots O(z_n,\bar{z}_n)}_{bc}$
can be expressed in terms of more and more complicated contour
integrals. The explicit perturbative computation of $\bvev{e^{-\int
    \lambda (z,\bar{z}) O(z,\bar{z}) \ud \bar{z}\wedge \ud z}}_{bc}$
looks even more formidable as it involves renormalization to remove
singularities in the $(z, \bar{z})$ integrals.  Anyway, many
interesting perturbations are not generically by minimal operators, as
shown be the example of the operator controlling the Hausdorff
dimension of SLE, which means perturbing the SLE measure using the
``natural'' length (i.e. the continuum limit of the discrete lattice
length) of interfaces.

One word on our strategy. That an interface measure is the result of
tracing over the other degrees of freedom of some statistical
mechanics model yields to some general compatibility conditions. At
criticality, this is the clue to relate SLE to conformal field theory
(CFT) : via the growth process construction of interfaces, CFT becomes
a provider of martingales for SLE, i.e. of observables which are
conserved in average under the growth process. Out of criticality, the
quantum field theory that describes macroscopic correlations in the
system close to criticality should for the same reasons be a
martingale provider for the corresponding interface measure. This is
hopefully enough to characterize this measure.  This is the approach
that we follow in this paper, using the ratio of partition functions
as observable.

SLE is most simply formulated in the upper half plane $\bH$.  There,
it describes curves originating from a boundary point that we choose
to be the origin $0$ of the real axis.  The curves $\gamma_{[0,t]}$,
parameterized by $t$, are coded in a conformal map $g_t$ uniformizing
$\bH\setminus\gamma_{[0,t]}$ onto $\bH$. To make this map unique we
require that its behavior at infinity is
$g_t(z)=z+{2t}/{z}+O(z^{-2})$. This is called the hydrodynamic
normalization with the parameter $t$ identified with half the
capacity.  The SLE measures are then defined by making the maps $g_t$
random and solutions of the stochastic Schramm-Loewner evolution:
$$
dg_t(z)=\frac{2dt}{g_t(z)-\xi_t},
$$
with $d\xi_t=\sqrt{\kappa}dB_t + F_t^0dt$ and $B_t$ a standard one
dimensional Brownian motion.  The points of the curves are
reconstructed from the maps $g_t$ via $\gamma_t=\lim_{\varepsilon\to
  0^+} g_t^{-1}(\xi_t+i\varepsilon)$ and the measure on the curves is
that induced via this reconstruction formula from the one on the maps
$g_t$.  Above $F^0_t$ is a possible drift which depends on the
variants of SLE one is considering. Different variants of SLE
correspond to different boundary conditions one imposes to the
critical statistical models.  As explained in \cite{BBKn, BBreport}
these drifts are intimately related to the partition functions of the
conformal field theories describing the continuum limit of these
statistical models.

Now look at perturbations away from criticality, with a perturbing
parameter $m$ which may depend on position. In the sequel we denote by
$P_m$ the corresponding measure on interfaces, so that $P_0$ is the
critical measure. We assume that the perturbation simply modifies the
drift (see the motivating discussion above, together with its
``caveat'') so that the Schramm-Loewner stochastic equation is
$$ d\xi_t=\sqrt{\kappa}dB^{[m]}_t+ F^{[m]}_tdt$$
with $B^{[m]}_t$ another standard Brownian motion and the drift
$F^{[m]}_t$ depending on the perturbation driving the systems out of
criticality.  However, contrary to the critical case, the off-critical
drift $F^{[m]}_t$ at 'time' $t$ depends on the full past history of
the curves\footnote{A word of caution is needed here. This phenomenon
  also happens for variants of SLE like SLE$_{\kappa,\rho ...}$ but in
  these cases one can introduce a finite number of auxiliary random
  processes in such a way as to get a usual (vector) Markov process.
  It is doubtful that such a trick exists for off-critical
  interfaces.}.

If the drift $F^{[m]}_t$ is well-defined, then under regularity
conditions, the off-critical measure can be shown to be regular with
respect to the critical one so that the Radon-Nikodym derivative $\ud
P_m/\ud P_0$ exists and the two measures differ by a density. In that
case, expectation values of events depending only on the curves up to
'time' $t$ differ in the off-critical $\EX^{[m]}[\cdots]$ and critical
$\EX[\cdots]$ measures by the insertion of a positive martingale:
$$ \EX^{[m]}[\cdots] = \EX[\, {\cal Z}_t^{[m]}\, \cdots ] $$
Here, ${\cal Z}_t^{[m]}$ has to be a positive martingale for the
critical process.  Its insertion reflects the difference between the
Boltzmann weights of the underlying statistical model at criticality
and away from it.  Again its existence is not guarantied. But in the
favorable case, by Girsanov's theorem \cite{Proba1,Proba2}, it is
linked to the off-critical drift by $F^{[m]}_t-F^0_t=\kappa\,
\partial_{\xi_t}\log {\cal Z}_t^{[m]}$.  The approach of
ref.\cite{BBK} relates ${\cal Z}_t^{[m]}$ to ratio of partition
functions of the quantum field theories describing the off-critical
models in the continuum limit.

Determining the martingale ${\cal Z}_t^{[m]}$ or the drift $F^{[m]}_t$
-- and proving that they make sense -- is a significant step towards
specifying what off-critical SLE is about. Of course it is only a
first step and a lot would remain to be done to determine and compute
properties of the off-critical curves.  One of the obstacles is that
we cannot rely on the Markov property of $\xi_t$ as in critical SLE.

\medskip

The aim of this paper is to determine ${\cal Z}_t^{[m]}$ and
$F^{[m]}_t$ in two simple cases: massive SLE(2) and massive SLE(4).

\medskip

Massive SLE(4) in its chordal version in $\bH$ describes curves from
$0$ to $\infty$ in the upper half plane.  Its corresponding field
theory is a massive Gaussian free field \footnote{We choose a position
  dependent mass so that all statements established here in the case
  of the upper half plane can be transported to any domain by
  conformal covariance. Under conformal transport by a map $g$ the
  mass is modified covariantly as $m(z)\to |g'(z)|\,m(z)$.}  which is
of course a non scale invariant perturbation -- by a mass term -- of
the free field conformal field theory associated to SLE(4).  We prove
that \footnote{Here and in the following, there is an implicit
  normalization constant to ensure that ${\cal Z}_{t=0}^{[m]}=1$.}
$$ {\cal Z}_t^{[m]}= \Big[
{\frac{{\rm Det}[-\Delta+m^2(z)]_{\bH_t}}{{\rm
      Det}[-\Delta]_{\bH_t}}}\Big]^{-\half}\, \exp{[-\int
  \frac{d^2z}{8\pi} m^2(z)\,\varphi_{t}(z) \Phi^{[m]}_{t}(z)]}
$$
is a (local) martingale for critical chordal SLE(4). Here the
determinants are determinants (regularized using $\zeta$-functions) of
the massive and massless Laplacian in the cut domain
$\bH_t\equiv\bH\setminus\gamma_{[0,t]}$ with Dirichlet boundary
conditions and $\varphi_{t}(z)$ and $\Phi^{[m]}_{t}(z)$ are the
one-point functions of the massless and massive free fields. They
satisfy $[-\Delta]\varphi_{t}=0$ and $[-\Delta+m^2]\Phi^{[m]}_{t}=0$
with appropriate discontinuous Dirichlet boundary conditions (with a
discontinuity of $\pi\sqrt{2}$ in our normalization).  The
off-critical drift for massive SLE(4) is:
$$ 
F^{[m]}_t = -\sqrt{2}\int \frac{d^2z}{2\pi} m^2(z)\
\Theta^{[m]}_{t}(z)\,\varphi_{t}(z) =-\sqrt{2}\int \frac{d^2z}{2\pi}
m^2(z)\ \theta_{t}(z)\,\Phi^{[m]}_{t}(z)
$$
with $\theta_{t}(z)$ and $\Theta^{[m]}_{t}(z)$ the massless and
massive Poisson kernel. See Section \ref{secmass} for details. For
this case, we have a satisfactory argument that the drift is indeed of
locally finite variation so that we are on the safe side of standard
stochastic calculus. In particular, $F^{[m]}_t$ is always nonnegative.

\begin{figure}
  \includegraphics[width=0.9\textwidth]{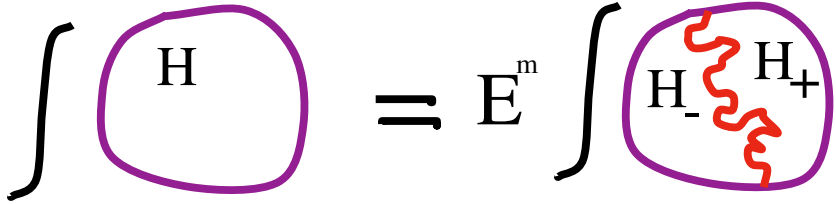}
  \vskip -2.0 truecm
  \caption{\emph{Decomposition of the Gaussian measure as the product
      of two Gaussian measures defined on each of the curves times the
      massive SLE measure on the curves.}}
  \label{fig:slematch}
\end{figure}

This drift can also be found by demanding that the one point function
$\Phi^{[m]}_{t}(z)$ is a martingale \cite{MS}.  Let $X$ be a Gaussian
free field with discontinuous Dirichlet boundary condition: $X=0$ on
$\bR_+$ and $X=\pi\lambda_c$ on $\bR_-$ (with $\lambda_c=\sqrt{2}$ in
our normalization).  We actually prove that any correlation function
of $X$ in the cut domain $\bH_t$, with an arbitrary number of marked
points, is a local martingale for massive SLE(4). Pushing this result
in the limit $t\to\infty$ provides arguments for the decomposition of
$X$ as the sum of two independent Gaussian fields. Namely, at infinite
time the curve $\gamma_{[0,\infty)}$ almost surely reaches the
boundary point at infinity \footnote{Here, we assume that this result
  proved in \cite{RS} for the critical SLE remains valid for massive
  SLE(4).} and it separates the domain $\bH$ in two sub-domains
$\bH_+$ and $\bH_-$ with $\bR_\pm$ part of the boundary of $\bH_\pm$.
Conditioned on $\gamma_{[0,\infty)}$, the field $X$ can be written as
the sum
$$ X=X_++X_-,\quad {\rm with}\quad
X_+\vert_{\partial\bH_+}=0,\quad
X_-\vert_{\partial\bH_-}=\pi\lambda_c,$$ where the fields $X_\pm$,
respectively restricted to $\bH_\pm$, are massive Gaussian free
fields. Consequently, conditioned on $\gamma_{[0,\infty)}$ the
Gaussian measure for $X$ can be factored as the product of the
Gaussian measures for $X_\pm$ so that:
\begin{eqnarray*}
  &&~~~~~~~~~~~~~
  \int_{X\vert_{\bR_+}=0}^{X\vert_{\bR_-}=\pi\lambda_c}
  \hskip -0.4truecm DX e^{-S_{m^2}[X]}\ [\cdots] \\
  &&  =\EX^{[m]}
  \int_{X_+\vert_{\partial\bH_+}=0}\hskip -0.8truecm DX_+ e^{-S_{m^2}[X_+]} 
  \int_{X_-\vert_{\partial\bH_-}=\pi\lambda_c} 
  \hskip -1.0truecm DX_- e^{-S_{m^2}[X_-]}\ [\cdots]
\end{eqnarray*}
for any observable $[\cdots]$. Here $S_{m^2}$ are the massive free
field actions and $\EX^{[m]}$ is the expectation with respect to the
massive SLE(4) measure. This decomposition strongly indicates that the
curve $\gamma_{[0,\infty)}$ may be seen as the discontinuity curve of
$X$, as proved in ref.\cite{SS} in the critical case.  See figure
(\ref{fig:slematch}).

Note that the expectation of the free field, $\vev{X(z)}^{[m]} \equiv
\Phi^{[m]}_{0}(z)$ has a simple interpretation as $\pi \lambda_c$
times the probability that the interface passes to the right of point
$z$.

\medskip

SLE(2) is the continuum limit of critical loop erased random walks
(LERW) as proved in the seminal Schramm's paper \cite{Sch0}.  Massive
SLE(2) describes a deformation of LERW in which the fugacity attached
to the underlying random walks has been moved away from criticality.
See e.g.  ref.\cite{BBK} for a more detailed introduction.  Its
associated field theory is that of a pair of massive symplectic
fermions.  We prove that, for any two marked points $a$ and $b$ on the
real axis,
$$
\tilde{\cal Z}_t^{[m]}= \Big[ {\frac{{\rm
      Det}[-\Delta+m^2(z)]_{\bH_t}}{{\rm Det}[-\Delta]_{\bH_t}}}\Big]
\times \Gamma^{[m]}_{t,[a,b]}
$$
is a local martingale for critical chordal SLE(2). Here
$\Gamma^{[m]}_{t,[a,b]}$ is an appropriate limit of a massive Poisson
kernel, see eq.(\ref{gamma_tm}) and Appendix \ref{ito_gamma}. At
criticality, $\Gamma^{[0]}_{t,[a,b]}$ is the chordal SLE(2) martingale
which intertwines chordal and dipolar SLE(2) (with marked points $a$
and $b$). Hence, $\tilde{\cal Z}_t^{[m]}$ is the martingale
intertwining critical chordal SLE(2) and massive dipolar SLE(2); i.e.
it describes the massive deformation of dipolar SLE(2). The
corresponding drift is:
$$ F^{[m]}_{t,[a,b]}=2\, \partial_{\xi_t}\log \Gamma^{[m]}_{t,[a,b]}.$$
This drift can alternatively be determined by requiring that
correlation functions of the symplectic fermions are local
martingales.

The ratio $\frac{\Gamma^{[m]}_{0,[x,y]}}{\Gamma^{[m]}_{0,[a,b]}}$ is
nothing but the probability that massive LERW dipolar with respect to
$[a,b]$ exits in the sub-interval $[x,y]\subset [a,b]$.

\medskip

The paper is organized as follows. In Section \ref{basics} we recall
basic facts about variants of critical SLE and about the formulation
of off-critical SLE "a la Girsanov" following ref.\cite{BBK}. In
Section \ref{secmass} we study massive SLE(4). We first compute the
drift using perturbation theory. We then prove non perturbatively that
${\cal Z}_t^{[m]}$, defined above, is a chordal SLE(4) local
martingale and re-derive the drift this way. We also prove that any
correlation functions of the massive Gaussian field in the cut domain
are martingales for massive SLE(4) and use this to derive the
decomposition of $X$ mentioned above.  In Section \ref{mlerw} we use
massive symplectic fermions to compute the drift and we prove that
$\tilde{\cal Z}_t^{[m]}$ is a critical chordal SLE(2) local
martingale. We also check that correlation functions of symplectic
fermions are massive SLE(2) local martingales and this provides
another way to derive the off-critical drift. Appendices
\ref{appendix1} and \ref{appendix2} are devoted to details concerning
the computation of the Ito derivative of the determinants of the
massive and massless Laplacian regularized using $\zeta$-functions.

\section{SLE basics and notations}\label{basics}

\subsection{Chordal and dipolar SLEs}
Here we recall the (by now standard) definition of SLE
\cite{Sch0,SLElawler}.  We shall use two variants of SLE: chordal and
dipolar. The former describes curves in a (planar) domain $\bD$ from a
boundary point to another boundary point, the latter describes curves
in $\bD$ from a boundary point to a sub-arc of the boundary of $\bD$.
In the following we choose $\bD$ to be the upper half plane
$\bH=\{z\in \bC,\ y\equiv \im z >0\}$, but our statements may be
transported to any planar simply connected domain by conformal
covariance.  In SLE, random curves $\gamma_{[0,t]}$, parameterized by
$t>0$, are coded into the conformal map which uniformizes $\bH_t\equiv
\bH\setminus \gamma_{[0,t]}$ onto $\bH$.  \medskip

$\bullet$ Chordal SLE in $\bH$ from $0$ to $\infty$. The Loewner
equation is
$$ \frac{d}{dt} g_t(z)=\frac{2}{g_t(z)-\xi_t}$$
with initial condition $g_0(z)=z$ and $\xi_t=\sqrt{\kappa} B_t$ a
Brownian motion with variance $\kappa$. The solution exits up to a
time $t$ for $z\in \bH\setminus \gamma_{[0,t]}$. The points of the
curves are such that $g_t(\gamma_t)=\xi_t$. Furthermore, $g_t$ is the
unique conformal map from $\bH\setminus \gamma_{[0,t]}$ to $\bH$ with
the hydrodynamic normalization $g_t(z)=z+{\cal O}(z^{-1})$, so that
any property of $g_t$ reflects one of the curve $\gamma_{[0,t]}$.  In
particular, the measure on the curves is that induces by the Brownian
motion.  \medskip

$\bullet$ Dipolar SLE in $\bH$ from $0$ to $[a,b]$.  It is a
particularly symmetric case of SLE$(\kappa,\rho)$.  The Loewner
equation is
\begin{eqnarray*}
  \frac{d}{dt} g_t(z)=\frac{2}{g_t(z)-\xi_t}&,&\quad 
  d\xi_t=\sqrt{\kappa} dB_t +F^0_t(a,b)dt\\
  F^0_{t,[a,b]}&=&\frac{(6-\kappa)/2}{a_t-\xi_t}+
  \frac{(6-\kappa)/2}{b_t-\xi_t}\\ 
  \frac{da_t}{dt}=\frac{2}{a_t-\xi_t}&,&\ \frac{db_t}{dt}=\frac{2}{b_t-\xi_t},
\end{eqnarray*}
that is $a_t=g_t(a)$ and $b_t=g_t(b)$.  Dipolar SLE is defined up to
time $T$ where $T>0$ is the random stopping time such that
$\gamma_T\in [a,b]$, i.e. the process is stopped at the moment it
touches the interval $[a,b]$.

\subsection{Intertwining variants of SLEs}
Girsanov's theorem describes the way stochastic equations are modified
by insertions of martingale weights in the measure
\cite{Proba1,Proba2}.  It provides a way to intertwine stochastic
equations with different drift terms.  In the physics literature, this
may be coded into the Martin-Siggia-Rose path integral representation
of stochastic differential equations.

More precisely, let $B_t$ be a Brownian motion and $\EX[\cdots]$ the
corresponding expectation. Let $M_t$ be a positive martingale with
respect to $\EX[\cdots]$.  To be a martingale implies that the Ito
derivative of $M_t$ is proportional to $dB_t$, so that we can write
$M_t^{-1}dM_t=f_t\, dB_t$. Then Girsanov's theorem tells us that with
respect to the weighted measure $\widehat \EX[\cdots]=\EX[M_T\
\cdots]$, the process $B_t$, $t<T$, satisfies the stochastic
differential equation $$dB_t=d\hat B_t + f_t\,dt$$ where $\hat B_t$ is
a Brownian motion with respect to $\widehat\EX[\cdots]$.  In other
word, inserting a martingale adds a drift in the stochastic equation
and reciprocally.

As an illustration, let us apply Girsanov's theorem to intertwine from
chordal SLE from $0$ to $\infty$ to dipolar SLE from $0$ to $[a,b]$.
 From the CFT/SLE correspondence \cite{BBsle}, martingales of chordal
SLE from $0$ to $\infty$ on $\bH$ may be constructed as CFT
correlation functions $\vev{{\cal
    O}\psi(\gamma_t)}_{\bH_t}/\vev{\psi(\infty)\psi(\gamma_t}_{\bH_t}$
with $\psi$ the operator (with scaling dimension $(6-\kappa)/2\kappa$)
creating the curve and ${\cal O}$ any spectator operator. To go from
chordal to dipolar SLE we need to choose ${\cal
  O}=\psi_{0;1/2}(a)\psi_{0;1/2}(b)$ with $\psi_{0;1/2}$ a primary
operator of dimension $h_{0;1/2}=(\kappa-2)(6-\kappa)/16\kappa$.  The
result is the following chordal SLE martingale
$$ \Gamma^0_{t,[a,b]}=|g_t'(a)g_t'(b)|^{h_{0;1/2}}\
|b_t-a_t|^{\frac{(\kappa-6)^2}{8\kappa}}
|\xi_t-a_t|^{\frac{(\kappa-6)}{2\kappa}}
|\xi_t-b_t|^{\frac{(\kappa-6)}{2\kappa}}$$ Its Ito derivative
reproduces the dipolar drift:
$$\sqrt{\kappa}\ {\Gamma^0_{t,[a,b]}}^{-1}(d\Gamma^0_{t,[a,b]}/dB_t)=
F^0_{t,[a,b]}=\frac{(6-\kappa)/2}{a_t-\xi_t}+\frac{(6-\kappa)/2}{b_t-\xi_t}$$
This is simply found by computing the logarithmic derivative of
$\Gamma^0_{t,[a,b]}$ with respect to $\xi_t=\sqrt{\kappa}\, B_t$.

\subsection{Off-critical SLEs}
We shall formulate off-critical SLE using the approach described in
ref.\cite{BBK} in which off-critical SLE is viewed as SLE twisted "a
la Girsanov" by a martingale, which we denote by ${\cal Z}_t^{[m]}$.
The off-critical measure is then $\EX^{[m]}[\dots] = \EX_{\rm
  SLE}[{\cal Z}_t^{[m]}~ \dots]$ so that the insertion of the
martingale ${\cal Z}_t^{[m]}$ amounts to weight differently SLE
configurations in a way reflecting the off-critical Boltzmann weights.
The off-critical martingales are ratio of partition functions
\footnote{As discussed in \cite{BBK}, there may also be an extra term
  in the formula for ${\cal Z}_t^{[m]}$ corresponding to a surface
  energy associated to the interface. But we do not need to include it
  at this point of the discussion.}:
$$ {\cal Z}_t^{[m]} = \frac{ \widehat Z_{\bH_t}^{[m]}}{\widehat
  Z_{\bH_t}^{[m=0]}}$$ where $\widehat Z_{\bH_t}^{[m]}=
Z_{\bH_t}^{[m]}/Z_{\bH}^{[m]}$ is the partition function of the
off-critical model (for $m\not= 0$ but critical for $m=0$) in the cut
domain normalized by that in the upper half plane.  See ref.\cite{BBK}
for a more detailed introduction and for extra (lattice) motivations.

Computing these martingales by taking the scaling limit of the
off-critical lattice model is an impossible task. In the continuous
field theory they may naively be presented as expectation values
$$ Z_{\bH_t}^{[m]}= \vev{ \exp\Big[-\int_{\bH_t}d^2z\, m^2(z)\Phi(z)\Big]\
  ({\rm "b.c.")}}_{\bH_t}$$ where the brackets $\vev{\cdots}$ refer to
critical CFT expectation values and the boundary conditions (``b.c.'')
are implemented by insertions of appropriate operators including the
operators generating the curves.  Of course this definition is plagued
with infinities and needs regularization and renormalization. As a
consequence of these infinities and of the fact that the perturbing
weight $ \exp[-\int_{\bH_t}d^2z\, m^2(z)\Phi(z)]$ is not a local
operator, it may turn out that ${\cal Z}_t^{[m]}$ is not a SLE
martingale although it is naively expected to be one since it is an
appropriate ratio of expectation values of CFT operators. See the
relevant discussion for self-avoiding-walks in \cite{BBK}.

One of the main aims and results of the following sections is to give
a precise meaning to ${\cal Z}_t^{[m]}$ in the case of massive SLE(4)
and SLE(2) and to prove that they are (local) martingales.

Assuming that ${\cal Z}_t^{[m]}$ is a martingale, Girsanov's theorem
tells us that the driving source in the Loewner equation satisfies the
stochastic equation
\begin{eqnarray}
  d\xi_t = \sqrt{\kappa} dB_t^{[m]} +  F^{[m]}_t dt,\quad {\rm with}\quad
  \sqrt{\kappa}\,{\cal Z}_t^{[m]\,-1}d{\cal Z}_t^{[m]}=(F^{[m]}_t-F_t^0)dB_t
  \label{girsamass}
\end{eqnarray}
with $B_t^{[m]}$ a Brownian motion with respect to $\EX^{[m]}[\dots]$
and $F^0_t$ the critical SLE drift.

In summary, off-critical SLEs may be defined using an appropriate
martingale ${\cal Z}_t^{[m]}$, provided that ${\cal Z}_t^{[m]}$ is
well-defined.  (This is not always the case as for instance in near
critical percolation \cite{NW}). Proving that it is a (local)
martingale amounts to show that the drift term in its Ito derivative
vanishes.  The drift term in the off-critical stochastic Loewner
equation is then given by $\sqrt{\kappa}{\cal Z}_t^{[m]\,-1}\,d{\cal
  Z}_t^{[m]}$.

\section{Massive SLE(4)}\label{secmass}
We look at massive SLE(4) in the chordal setting describing curves
from $0$ to $\infty$ in $\bH$.  As shown by Sheffield and Schramm
\cite{SS}, samples of SLE(4) may be viewed as discontinuity lines of
samples of a Gaussian massless free field.  The aim of this section is
to describe what happens to these lines when we consider a massive
Gaussian free field.

\subsection{SLE(4) and free massless boson}
A Gaussian massless free field is a conformal field theory with
central charge $c=1$.  Denoting by $X$ the free field, its action is:
$$ S_0[X]= \int \frac{d^2z}{2\pi}\ (\partial X)(z)(\bar \partial X)(z)$$
with $d^2z$ the Lebesgue measure.  For simplicity we first consider
the system in the upper half plane $\bH$ \footnote{Points in the
  complex plane will be identified to complex numbers $z=x+iy,\ \bar
  z=x-iy$ with $(x,y)$ real, $y>0$.  We denote
  $\partial=\half(\partial_x-i\partial_y)$ and $\bar
  \partial=\half(\partial_x+i\partial_y)$.  The Laplacian is
  $\Delta=4\partial \bar \partial$.}, but we may extend our discussion
to any domain by conformal covariance.  We impose Dirichlet boundary
conditions on the real axis $\bR$ with a discontinuity at the origin,
so that $X\vert_{\bR_+}$ and $X\vert_{\bR_-}$ are constants on the
positive real axis and on the negative real axis respectively.  The
discontinuity at $0$, $X\vert_{\bR_-}-X\vert_{\bR_+}$ is written as
$\pi\lambda_c$ and the constant $\lambda_c$ will be fixed to the
critical value $\lambda_c= \sqrt{2}$ to ensure a perfect matching
between chordal SLE(4) from $0$ to $\infty$ and the Gaussian massless
free field.

Let us note that in the massless (critical) theory the symmetry $X
\rightarrow X+ c^{st}$ implies that only the value of the
discontinuity at $0$ matters, not the individual constants
$X\vert_{\bR_+}$ and $X\vert_{\bR_-}$. This is not true when the
perturbation is turned on and changing $X\vert_{\bR_+}$ for fixed
$\lambda_c$ changes the statistics of the interface. For compactness
what follows is written assuming that $X\vert_{\bR_+}=0$, but all
formul\ae\ below remain correct in the general case if the appropriate
one point (massive and massless) functions are used. The connected two
point functions are not affected by a translation of the boundary
conditions.

With the boundary conditions, $X\vert_{\bR_+}=0$ and
$X\vert_{\bR_-}=\pi\lambda_c$ the massless one and two point functions
are:
\begin{eqnarray*}
  \varphi_\bH(z)&\equiv& \vev{X(z)}_\bH=\lambda_c\ \im \log z,\\
  G_\bH(z,w)&\equiv& \vev{X(z)X(w)}^c_\bH
  =-\log\big\vert\frac{z-w}{z-\bar w}\big\vert^2
\end{eqnarray*}
where $\vev{X(z)X(w)}^c$ denotes the connected two-point function,
defined as $\vev{X(z)X(w)}^c=\vev{X(z)X(w)} -\vev{X(z)}\vev{X(w)}$.
Here $G_\bH$ is the Green function of the Laplacian with Dirichlet
boundary conditions: $-\Delta G_\bH(z,w)=4\pi \delta^{(2)}(z,w)$ with
$\delta^{(2)}(\cdot,\cdot)$ the Dirac point measure.

In a maybe more probabilistic verbatim, $X$ may be viewed as a
Gaussian distribution valued variable with characteristic function:
$$\vev{e^{(J,X)}}_\bH=
\exp{[\, \int d^2z\, J(z)\varphi_\bH(z) +\half \int d^2zd^2w\,
  J(z)G_\bH(z,w)J(w)\, ]}$$ for any source $J(z)$ suitably
well-behaved on the upper half plane and $(J,X)=\int d^2z\, J(z)X(z)$.

To couple this Gaussian massless free field to SLE(4) we consider its
correlation functions in the domain $\bH_t$ cut along a SLE sample:
$\bH_t\equiv \bH\setminus \gamma_{[0,t]}$.  Since $X$ is a scalar
field, its expectation values in $\bH_t$ are simply computed from
those in $\bH$ by conformal transport. If $h_t(z)\equiv g_t(z)-2B_t$
denotes the uniformizing SLE(4) map from $\bH_t$ onto $\bH$ mapping
the tip of the curve back to the origin, $h_t(\gamma_t)=0$, we have:
\begin{eqnarray*}\
  \varphi_{t}(z)&\equiv& \vev{X(z)}_{\bH_t} = \varphi_{\bH}(h_t(z)),\\
  G_{t}(z,w)&\equiv& \vev{X(z)X(w)}^c_{\bH_t} =G_{\bH}(h_t(z),h_t(w))
\end{eqnarray*}

As known from the SLE/CFT correspondence \cite{BBsle}, multi-point
correlation functions of the Gaussian massless free field in the cut
domain are SLE(4) (local) martingales.  This is true for the one-point
function, as it can be checked by computing its Ito derivative,
\begin{equation}\label{dvarphi}
  d \varphi_{t}(z) = \lambda_c\, \theta_{t}(z)\, dB_t,\quad
  \theta_{t}(z)\equiv - \im \frac{2}{h_t(z)},
\end{equation}
but also for the non connected two-point function iff $\lambda_c^2=2$,
as it follows from the Hadamard formula which gives the variation of
the Green function:
$$ d G_{t}(z,w) = -2\,  \theta_{t}(z)\theta_{t}(w)\, dt$$
As a consequence since the theory is Gaussian, this is also true for
the characteristic function for any source $J$ but in the cut domain
$\bH_t$, so that
\begin{eqnarray}
  \vev{e^{(J,X)}}_{\bH_t}\quad {\rm is\ an\ SLE(4)\ martingale}.
  \label{critmatch}
\end{eqnarray}
All multi-point correlation functions of $X$ in the cut domain are
discontinuous along $\gamma_{[0,t]}$ with a jump of $\lambda_c$
indicating that effectively $\gamma_{[0,t]}$ is almost surely the
discontinuity lines of $X$.  Notice that this requires adjusting the
Dirichlet discontinuity to its critical value $\lambda_c=\sqrt{2}$.

\subsection{Massive perturbation}
We consider perturbing the massless action by a mass term:
$$ S_{m^2}[X]= \int \frac{d^2z}{8\pi} \big[ 4(\partial X)(z)(\bar
\partial X)(z) + m^2(z)X^2(z) \big] $$ We assume the mass to be
position dependent in order to avoid possible infra-red (large
distance) divergences and also to make the theory conformally
covariant (but at the prize of modifying the mass when implementing
conformal transformations). As before we consider the theory on
$\bH_t$ and write the correlation functions in the massive theory with
the mass as an upper index, e.g.  $\langle \sO \rangle^{[m]}_{\bH_t}$.
We impose discontinuous Dirichlet boundary conditions as in the
massless case. Namely: $X=0$ and $X=\pi\lambda_c$, respectively to the
right and to the left of the tip of the curve $\gamma_t$.

With this definition, the one-point function in the massive theory is:
$$ \vev{X(z)}^{[m]}_{\bH_t}= \Phi^{[m]}_t(z)$$
with $\Phi^{[m]}_t(z)$ the solution of the classical equation of
motion $[-\Delta+m^2(z)]\Phi^{[m]}_t(z)=0$ with discontinuous
Dirichlet boundary conditions as defined above.  The connected
two-point function is the massive Green function:
$$\vev{X(z)X(w)}^{[m];c}_{\bH_t}=G^{[m]}_t(z,w)$$
with $[-\Delta+m^2(z)]G^{[m]}_t(z,w)=4\pi\delta^{(2)}(z,w)$ and
$G^{[m]}_t(z,w)=0$ for $z$ or $w$ on the boundary of $\bH_t$.

Alternatively, the massive Gaussian free field may be defined by its
generating functions:
$$\vev{e^{(J,X)}}^{[m]}_{\bH_t}=
\exp\left[\, \int d^2z\, J(z)\Phi^{[m]}_t(z) +\half \int d^2zd^2w\,
  J(z)G^{[m]}_t(z,w)J(w)\, \right]$$ for any source $J(z)$.

An explicit expression for the massive classical solution
$\Phi^{[m]}_t(z)$ may be written in terms of the massless solution and
the massive Green function:
$$\Phi^{[m]}_t(z)= \varphi_t(z)-
\frac{1}{4\pi} G^{[m]}_t(z,\cdot)\star m^2(\cdot)\varphi_t(\cdot)$$
where $\star$ denotes the convolution product~\footnote{The
  convolution is defined in the usual way $H(z,\cdot)\star f(\cdot)=
  \int d^2z'\,H(z,z')\,f(z') $}.  For later convenience we also need
to introduce the so-called massive Poisson kernel defined similarly
as:
$$\Theta^{[m]}_t(z)= \theta_t(z)-
\frac{1}{4\pi} G^{[m]}_t(z,\cdot)\star m^2(\cdot)\theta_t(\cdot)$$ Of
course the massive Green function satisfies a convolution formula
whose iteration reproduces the perturbative series.

\subsection{The off-critical drift}\label{OCDrift}
Recall from Girsanov's theorem that the off-critical drift $F^{[m]}_t$
at $\kappa=4$ is given by the Ito derivative of the partition function
martingale: $F^{[m]}_t\, dB_t= 2{\cal Z}_t^{[m]\, -1}\, d{\cal
  Z}_t^{[m]}$ with ${\cal Z}_t^{[m]}$ the massive partition function
(normalized by the massless one so that ${\cal Z}_t^{[m=0]}=1$) in the
cut upper half plane~\footnote{Here we assume (and we shall prove it
  in the following) that ${\cal Z}_t^{[m]}$ is a SLE(4) martingale.}:
$$ {\cal Z}_t^{[m]} \equiv \vev{\exp[ - \int\frac{d^2z}{8\pi}\
  m^2(z)X^2(z) \,]}_{\bH_t}$$ where the expectation is with respect to
the massless Gaussian measure.

We have to give a meaning to $X^2$. This is done via a point splitting
subtraction of the logarithmic singularity in $X(z)X(w)$ as $w$
approaches $z$:
\begin{eqnarray}
  X^2(z) \equiv \lim_{w\to z} X(z)X(w) +\log\vert z-w\vert^2
  \label{pointsplit}
\end{eqnarray}
It is a local definition and insertions of $X^2$ are then well-behaved
in any expectation values.  With this definition, ${\cal Z}_t^{[m]}$
is finite (in any order in perturbation theory).

\subsubsection{First order computation}
To first order in perturbation theory, the massive partition function
is:
$$ {\cal Z}_t^{[m]} = 1- \int\frac{d^2z}{8\pi}
m^2(z)\vev{X^2(z)}_{\bH_t}+\cdots$$ Although $X$ is a scalar -- and
thus it transforms as a scalar under conformal transformations --,
$X^2$ is not a scalar as the logarithmic subtraction in the point
splitting definition produces an anomaly in its transformation laws.
As a consequence its one-point function in the cut domain $\bH_t$ is:
$$\vev{X^2(z)}_{\bH_t}= \varphi_{t}^2(z) + 2\log \rho_{t}(z)$$
where $\rho_{t}(z)\equiv {2\im h_t(z)}/{\vert h_t'(z)\vert}$ is the
conformal radius at $z$ which, by Kobe's theorem, is an estimate of
the distance between $z$ and the boundary of $\bH_t$.

The formula for $\vev{X^2(z)}_{\bH_t}$ has a nice probabilistic
interpretation.  By construction, $\varphi_{t}(z)$ is a SLE(4)
martingale (recall that $d \varphi_{t}(z) = \lambda_c\,
\theta_{t}(z)\, dB_t$), but its square is not. However, as a CFT
expectation value in $\bH_t$, $\vev{X^2(z)}_{\bH_t}$ is a SLE(4)
martingale.  So, $2\,\log \rho_{t}(z)$ is what is needed to be added
to $\varphi_{t}^2(z)$ to make it a martingale, i.e. its times
derivative is the quadratic variation of $\varphi_{t}(z)$, provided
(again) that $\lambda^2_c=2$.  Explicitly $d\log \rho_{t}(z)=-(\im
2/h_t(z))^2\, dt$. As a consequence:
$$ d \vev{X^2(z)}_{\bH_t}= 2 \varphi_{t}(z) d\varphi_{t}(z)
= 2\lambda_c\, \theta_{t}(z)\, \varphi_{t}(z)\, dB_t$$

Computing the off-critical drift to first order is now very easy. We
just have to Ito differentiate the partition function and, permuting
integration and Ito derivative\footnote{There is no problem in doing
  this permutation as the integrand is regular enough.}, we get:
$${\cal Z}_t^{[m]\, -1}\, d{\cal Z}_t^{[m]}
= -2\lambda_c\, \int\frac{d^2z}{8\pi} m^2(z)\, \theta_{t}(z)\,
\varphi_{t}(z) dB_t+\cdots$$ where the dots refer to higher order term
in the mass perturbation.

\subsubsection{All order computation}
Since the theory is Gaussian the partition function ${\cal Z}_t^{[m]}$
can be computed to all orders. Let us assume for a while that this
partition function is an SLE(4) martingale.  This will be proved in
the following section.  To determine the drift we need to compute
${\cal Z}_t^{[m]\, -1}\, d{\cal Z}_t^{[m]}$. Since we only have to
extract the term proportional to $dB_t$, which is a first order term
in the Ito derivative, (the higher order terms in the Ito derivative
would cancel as ${\cal Z}_t^{[m]}$ is a martingale), it is enough to
look at the first order $dB_t$ term in $\log {\cal Z}_t^{[m]}$.  In
perturbative expansion, $\log {\cal Z}_t^{[m]}$ is the sum of the
connected diagrams:
$$\log {\cal Z}_t^{[m]} = \sum_{n\geq 0} \frac{(-)^n}{n!}
\int \prod_{j=1}^n \frac{d^2z_j}{8\pi} m^2(z_j)\ \cdot\
\vev{X^2(z_1)\cdots X^2(z_n)}_{\bH_t}^{\rm connected}$$ There are two
types of connected diagrams: (i) diagrams which produce terms like
$G_{t}(z_1,z_2)\cdots G_{t}(z_{n-1},z_n) G_{t}(z_n,z_1)$ up to
permutations -- there are $2^{n-1}(n-1)!$ such diagrams -- and (ii)
diagrams which produce terms like
$\varphi_{t}(z_1)G_{t}(z_1,z_2)\cdots G_{t}(z_{n-1},z_n)
\varphi_{t}(z_n)$ up to permutations -- there are $2^{n-1}n!$ such
diagrams.  Only diagrams of the second kind contribute to the $dB_t$
term in the Ito derivative because the first ones only involve the
Green function.  Using $d \varphi_{t}(z)=\lambda_c\,\theta_{t}(z)dB_t$
and summing up, we find:
\begin{eqnarray*} {\cal Z}_t^{[m]\, -1}\, d{\cal Z}_t^{[m]}&=& dB_t\,
  \sum_{n\geq 1} (-2)^n \lambda_c\int \prod_{j=1}^n
  \frac{d^2z_j}{8\pi} m^2(z_j) \times\\
  &&\times\, \theta_{t}(z_1)\, G_{t}(z_1,z_2)\cdots
  G_{t}(z_{n-1},z_n)\, \varphi_{t}(z_n)
\end{eqnarray*}
The sum reproduces the perturbative expansion of the massive Green
function:
$$
-2\lambda_c \int \frac{d^2z}{8\pi} m^2(z)\ \theta_{t}(z)\,
\big[\varphi_{t}(z)-\frac{1}{4\pi} G^{[m]}_{t}(z,\cdot)\star
m^2(\cdot) \varphi_{t}(\cdot)\big]
$$
where again $\star$ denotes convolution. We here recognize the
solution of the classical equation of motion $\Phi^{[m]}_t$.  Thus
$(\lambda_c=\sqrt{2})$:
$${\cal Z}_t^{[m]\, -1}\, d{\cal Z}_t^{[m]}\,= -2\lambda_c
\int \frac{d^2z}{8\pi} m^2(z)\ \theta_{t}(z)\Phi^{[m]}_{t}(z)\ dB_t.
$$
Since $G^{[m]}_{t}$ is symmetric, we can also write the drift as:
\begin{eqnarray}
  F^{[m]}_t= -2\sqrt{2}\int \frac{d^2z}{4\pi} m^2(z)\  
  \Theta^{[m]}_{t}(z)\,\varphi_{t}(z)
  \label{drfit4}
\end{eqnarray}
Recall that $\sqrt{\kappa} {\cal Z}_t^{[m]\, -1}\, d{\cal
  Z}_t^{[m]}=F^{[m]}_t\,dB_t$.  In the next Sections we will see two
different ways of obtaining this result.

\subsection{Perfect matching and decomposition}
 From basic rules of statistical mechanics, we expect that massive
correlation functions in the cut domain are martingales for massive
SLEs.  This is how Makarov and Smirnov computed the off-critical drift
for massive SLE(4)~\footnote{We thank S. Smirnov for a discussion
  concerning this point.}.

Let us first look at the one-point function
$\vev{X(z)}^{[m]}_{\bH_t}$. This correlation function is the
probability that the massive SLE curve passes to the right of point
$z$, conditioned on the beginning of the curve up to time $t$. The
argument leading to this result is the same as in the massless case
and it uses the fact that $\vev{X(z)}^{[m]}_{\bH_t}$ is a martingale
for the massive SLE. In order to check this property recall that:
$$\vev{X(z)}^{[m]}_{\bH_t}=\Phi^{[m]}_{t}(z)
=\varphi_{t}(z)- \frac{1}{4\pi} G^{[m]}_{t}(z,\cdot)\star
m^2(\cdot)\varphi_{t}(\cdot)$$ with
$\varphi_{t}(z)=\varphi_{\bH}(h_t(z))$. Computing its Ito derivative
we have $d\varphi_{t}(z)= \lambda_c\theta_{t}(z)dB_t$.  Recall that
$2dB_t=2dB^{[m]}_t+F^{[m]}_tdt$.  To compute $d\Phi^{[m]}_{t}(z)$ we
need to known the derivative of the massive Green function. This is
provided by the massive Hadamard formula (which follows for instance
from the massless Hadamard formula and the convolution formula
satisfied by the Green function):
$$dG^{[m]}_{t}(z,w)= 
-2\, \Theta^{[m]}_{t}(z)\Theta^{[m]}_{t}(w)\, dt$$ This gives (with
$\lambda_c=\sqrt{2}$):
\begin{eqnarray*}
  d\Phi^{[m]}_{t}(z)&=& \lambda_c\, \Theta^{[m]}_{t}(z)\,
  [dB^{[m]}_t+\half F^{[m]}_tdt]\\
  && + \Theta^{[m]}_{t}(z) dt\cdot \int \frac{d^2w}{2\pi} m^2(w)\,
  \Theta^{[m]}_{t}(w)\varphi_{t}(w)
\end{eqnarray*}
Hence, $\Phi^{[m]}_{t}(z)$ is a $P_m$ local martingale provided the
drift is:
$$ F^{[m]}_t = - 2\sqrt{2}\int \frac{d^2w}{4\pi} m^2(w)\,
\Theta^{[m]}_{t}(w)\varphi_{t}(w)$$ which coincides with what we
field-theoretically computed in the previous section.  Notice that
then:
$$ d\Phi^{[m]}_{t}(z)= \lambda_c\, \Theta^{[m]}_{t}(z)\,dB^{[m]}_t$$
Consider now the two-point function $\vev{X(z)X(w)}^{[m]}_{\bH_t}$
which is the sum of the product of two one-point functions plus the
massive Green function. Thanks to the massive Hadamard formula and to
the formula for $d\Phi^{[m]}_{t}(z)$ it is then readily checked that
$\vev{X(z)X(w)}^{[m]}_{\bH_t}$ is a martingale (i.e.  the drift term
vanishes) provided that $\lambda_c^2=2$.

Since the theory is Gaussian, the fact that the one and two point
functions are martingales implies that any $n$-point function is a
local martingale.  This is also true for the generating function:
\begin{eqnarray} \label{fullmartin} \vev{e^{(J,X)}}^{[m]}_{\bH_t}\quad
  {\rm is\ a} \quad P_m {\rm -SLE(4)\ martingale}
\end{eqnarray}
for any source $J$ (with compact support, say).  This was expected
from naive statistical mechanics arguments.  Statement
(\ref{fullmartin}) actually needs a few justifications because it
applies to the exponential of the integral of a martingale.  Consider
first the integrated one-point function $I_t\equiv\int d^2z
J(z)\Phi^{[m]}_{t}(z)$. We know that $\Phi^{[m]}_{t}(z)$ is a bounded
local martingale and thus a martingale. It is also positive. Thus
Fubini's theorem applies and we can permute the $d^2z$ integration and
the expectation $\EX^{[m]}$ which is enough to prove that $I_t$ is a
bounded martingale.  Consider now the integrated two-point functions.
$I_t^2$ is not a martingale but $I^2_t$ minus its quadratic variation
$(\delta I_t)^2$ is a martingale \cite{Proba1,Proba2}. This quadratic
variation is bilinear in the current $J$.  Considering $J$'s equal to
a sum a weighted Dirac measures localized at arbitrary points then
determined this bilinear form and $(\delta I_t)^2=-\int
d^2zd^2wJ(z)\,\Delta G^{[m]}_t(z,w)\,J(w)$ with $\Delta
G^{[m]}_t\equiv G^{[m]}_t-G^{[m]}_0$. Finally, the exponential
$\vev{e^{(J,X)}}^{[m]}_{\bH_t}=e^{I_t-\half(\delta I_t)^2}$ is a
bounded local martingale and thus a martingale.

We now use the property (\ref{fullmartin}) to derive the decomposition
of $X$ mentioned in the introduction. In the limit $t\to\infty$, this
property gives that
$$ \EX^{[m]}[\, \vev{e^{(J,X)}}^{[m]}_{\bH_\infty}\,]= 
\vev{e^{(J,X)}}^{[m]}_{\bH}$$ where $\EX^{[m]}$ is the massive SLE(4)
measure on the complete curve $\gamma_{[0,\infty)}$. Almost surely
(this was proved in the critical case but we assumed it is still true
in the massive case), the curve $\gamma_{[0,\infty)}$ reaches infinity
and cuts the domain $\bH$ in two part $\bH_+$ and $\bH_-$ whose
boundaries are respectively $\bR_+$ (or $\bR_-$) and the right
$\gamma_{[0,\infty)}^+$ (or the left $\gamma_{[0,\infty)}^-$) side of
the curve.  The expectations $\vev{e^{(J,X)}}^{[m]}_{\bH_\infty}$ are
fully determined by the limiting behavior as $t\to\infty$ of the one
and two point functions. Almost surely, we have:
\begin{eqnarray*}
  \lim_{t\to\infty} \Phi^{[m]}_{t}(z)=
  \begin{cases}
    0,& z\in\bH_+ \\ \pi\sqrt{2}, & z\in\bH_-
  \end{cases}
\end{eqnarray*}
and
\begin{eqnarray*}
  \lim_{t\to\infty} G^{[m]}_{t}(z,w)=
  \begin{cases}
    G^{[m]}_{\bH_-}(z,w),& z,w\in\bH_-  \\
    0,& z\in\bH_-,\ w\in\bH_+\\
    G^{[m]}_{\bH_+}(z,w), & z,w\in\bH_+
  \end{cases}
\end{eqnarray*}		
where $G^{[m]}_{\bH_\pm}$ are the massive Green functions in the two
sub-domains $\bH_\pm$ with Dirichlet boundary conditions.  If these
limits exist their values can only be those written above because of
the differential equations they satisfy. So we only have to argue that
they exist. In the massless case, convergence of the one-point
function was proved in \cite{SchPerco} based on the fact that
$\varphi_t(z)$ is proportional to the harmonic measure of
$\bR_-\cup\gamma_{[0,t]}^-$ viewed from $z$.  Convergence of the
massless Green function is based on the fact that $G^{[0]}_t$ and
$G^{[0]}_{\bH_\pm}$ are solutions of the same differential equations
with slightly different boundary conditions but whose difference
converges to zero as $t\to\infty$. Let us sketch the argument. Assume
for instance that $z,w\in \bH_+$ and consider the differences
$G^{[0]}_t-G^{[0]}_{\bH}$ and $G^{[0]}_{\bH_+}-G^{[0]}_{\bH}$, say as
functions of $z$ at $w$ fixed.  The first one is harmonic on $\bH_t$,
it reaches its maximum on the boundary $\partial\bH_t$ and this
maximum is bounded by ${\rm max}_{\gamma_{[0,\infty)}}G^{[0]}_{\bH}$.
The second one is harmonic on $\bH_+$, it reaches its maximum on the
boundary $\partial\bH_+$ which is therefore also bounded by ${\rm
  max}_{\gamma_{[0,\infty)}}G^{[0]}_{\bH}$.  Hence, the difference
$G^{[0]}_t-G^{[0]}_{\bH_+}$ is harmonic on $\bH_+$, with boundary
condition bounded by $2{\rm max}_{\gamma_{[0,\infty)}}G^{[0]}_{\bH}$
and non-vanishing only a sub-arc of the boundary of the domain
vanishing as $t\to\infty$ (because almost surely the curve
$\gamma_{[0,t)}$ goes to infinity). Similar arguments apply for
$z\in\bH_+$ and $w\in\bH_-$.The functional relations satisfied by the
massive and the massless Green functions and the one-point functions
imply that once the statement is proved for the massless quantities it
is also true for the massive one.

As a consequence $\vev{e^{(J,X)}}^{[m]}_{\bH_\infty}$ factors into the
product of expectations in the two sub-domains, as expected:
$$ \vev{e^{(J,X)}}^{[m]}_{\bH_\infty}
=\vev{e^{(J,X)}}^{[m]}_{\bH_+}\times \vev{e^{(J,X)}}^{[m]}_{\bH_-}$$
In each sub-domains, correlation functions are those of a Gaussian
free field with Dirichlet boundary conditions $0$ in $\bH_+$ and
$\pi\sqrt{2}$ in $\bH_-$. That is: conditioned on
$\gamma_{[0,\infty)}$ the field $X$ can be decomposed as the sum
$X=X_++X_-$ of two Gaussian fields $X_\pm$ respectively defined on
$\bH_\pm$ with Dirichlet boundary conditions ($0$ in $\bH_+$ and
$\pi\sqrt{2}$ in $\bH_-$), as mentioned in the introduction.

\subsection{Partition functions and the off-critical martingale}
\label{sle4mart}
We have seen in Section \ref{OCDrift} that as a consequence of
Girsanov's theorem we can compute the off-critical drift by taking the
Ito derivative of the ratio of massive and massless partition
functions with discontinuous Dirichlet boundary conditions. This can
be written as a correlation function in the massless theory
$$
{\cal Z}_t^{[m]} = \vev{\exp[ - \int\frac{d^2z}{8\pi}\ m^2(z)X^2(z)
  \,]}_{\bH_t}
$$

The usual heuristic arguments from statistical mechanics tell us that
this is a martingale for the critical SLE.  Actually, since both the
massless and the massive theories are Gaussian, one can compute their
partition functions in a fully non perturbative way. This allows us to
prove rigorously that the ratio of the massive/massless partition
functions is a (local) martingale for the critical measure and at the
same time to compute the off-critical drift.  The simplest way to
proceed is by first decomposing $X$ as the sum of its one-point
function plus a Gaussian field $\bar X$ with zero Dirichlet boundary
conditions.  In the cut domain $\bH_t$ this reads: $X=
\varphi_{t}+\bar X$.  Notice that this decomposition is done on the
massless Gaussian field as the partition function is defined via an
expectation value in the massless theory.  Then $X^2= \varphi_{t}^2+2
\varphi_{t}\bar X + \bar X^2$ (with $\bar X^2$ defined with a similar
point splitting regularization) and the expectation value can be
reduced to an expectation value in the boundary zero Gaussian field.
Thus:
$${\cal Z}_t^{[m]}= {\cal Z}_t^{[m];\bar X}\,\cdot
e^{-\int \frac{d^2z}{8\pi} m^2(z)\,\varphi_{t}^2(z)}\, \vev{e^{-2\int
    \frac{d^2z}{8\pi} m^2(z)\,\varphi_{t}(z)\, \bar
    X(z)}}^{[m]}_{\bH_t}$$ Here ${\cal Z}_t^{[m];\bar X}$ is the
partition function (relative to the massless theory) of the massive
boundary zero Gaussian field and the last expectation value is an
expectation value in the massive boundary zero Gaussian field. It is
thus equal to
$$\exp\left[ \half \int \frac{d^2z}{4\pi}\frac{d^2w}{4\pi}\, 
  m^2(z)\varphi_{t}(z)G^{[m]}_{t}(z,w)m^2(w)\varphi_{t}(w)\, \right]$$
The integration over $w$ involves the convolution of
$G^{[m]}_{t}(z,\cdot)$ with $m^2(\cdot)\varphi_{t}(\cdot)$ which,
combined with the function $\varphi_{t}(z)$ in the second factor of
the previous expression of the partition function, reproduces the
massive classical solution $\Phi^{[m]}_{t}(z)$.  Hence,
\begin{equation} {\cal Z}_t^{[m]}= {\cal Z}_t^{[m]; \bar X}\,\cdot
  \exp[-\int \frac{d^2z}{8\pi} m^2(z)\,\varphi_{t}(z)
  \Phi^{[m]}_{t}(z)]
\end{equation}
The partition function ${\cal Z}_t^{[m]; \bar X}$ is the ratio of the
square roots of the determinants of the massive and massless Laplacian
with Dirichlet boundary conditions:
$$ {\cal Z}_t^{[m];\bar X}= \Big[{
  \frac{{\rm Det}[-\Delta+m^2(z)]_{\bH_t}}{{\rm Det}[-\Delta]_{\bH_t}}
}\Big]^{-\half}
$$

\subsection{A representation of the partition function}
To arrive at an alternative representation of the partition function,
let us introduce a fictitious parameter $\tau$ multiplying $m^2(z)$,
and consider the path integral representation of the determinant of
the massive Laplacian with Dirichlet boundary conditions. Taking the
derivative with respect to $\tau$ we get
\begin{equation*}
  \frac{d}{d\tau} {\rm Det}[-\Delta+ \tau
  m^2(z)]_{\bH_t}^{-\half}=-    
  \int{\cal D}\bar X
  \left(\int\frac{d^2z}{8\pi}m^2(z)\,\bar X^2(z)\right) 
  e^{-S_{\tau m^2}[\bar X]} 
\end{equation*}
Hence
\begin{equation}\label{dlogZtau}
  \frac{d}{d\tau}  \log\,{\cal Z}_t^{[\tau m];\bar X} =-  
  \int\frac{d^2z}{8\pi}m^2(z)\,\langle \bar
  X^2(z)\rangle^{[\sqrt{\tau}m]}_{\bH_t}  
\end{equation}

Of course this result is only formal. We have given no prescription
how to regularize the composite operator $\bar X^2(z)$. The proper
computation, which is done in appendix \ref{appendix1}, uses the
definition of the functional determinant through the $\zeta$-function
regularization.  It turns out that -- up to an irrelevant term
proportional to $\int \frac{d^2z}{4\pi} m^2(z)$ --, the
$\zeta$-function regularization corresponds to the point splitting
regularization of $\bar X^2(z)$ (as done in the perturbative
computation of Section \ref{OCDrift}, see eq.(\ref{pointsplit})):
$$
\langle \bar X^2(z)\rangle^{[\sqrt{\tau}m]}_{\bH_t} =
\lim_{z'\rightarrow z} \langle \bar X(z')\bar
X(z')\rangle^{[\sqrt{\tau}m]}_{\bH_t} +\log\vert z'-z\vert^2
$$
Integrating back eq.(\ref{dlogZtau}) and inserting the expression for
$\langle \bar X^2(z)\rangle^{[\sqrt{\tau}m]}_{\bH_t}$ we arrive at:
\begin{equation}\label{log1}
  \log {\cal Z}_t^{[m];\bar X} = -\int\frac{d^2z}{8\pi}
  m^2(z)\left[ \log|\rho_{t}(z)|^2+ \int_0^1 
    K_{t}^{[\sqrt{\tau} m]}(z)d\tau \right] 
\end{equation}
where
\begin{equation}\label{Ktilde}
  K_{t}^{[m]}(z) \equiv \lim_{z'\rightarrow z}
  (G^{[m]}_{t}- G^{[0]}_{t})(z',z)=
  -\int\frac{d^2z'}{4\pi}G^{[m]}_{t}(z,z')m^2(z')G^{[0]}_{t}(z',z),
\end{equation}
and the integrals are convergent.

\subsubsection{Proof that ${\cal Z}_t^{[m]}$ is a martingale}
To prove that ${\cal Z}_t^{[m]}$ is a local martingale, we use its
representation in eq.(\ref{log1}) and compute its Ito derivative.
Evaluating separately the Ito derivative of ${\cal Z}_t^{[m];\bar X}$
and of $\exp[-\int \frac{d^2z}{8\pi} m^2(z)\,\varphi_{t}(z)
\Phi^{[m]}_{t}(z)]$ would lead to the appearance of diverging
integrals. In order to avoid this problem we perform a slightly
different splitting by extracting the logarithm of the conformal
radius from formula (\ref{log1}) and putting it together with ${\cal
  Z}_t^{[m];\bar X}$. We therefore write
$${\cal Z}_t^{[m]}={\tilde {\cal Z}}_t^{[m];\bar X}\ Y_t$$ 
where we have defined
$$ Y_t \equiv \exp\left[-\int
  \frac{d^2z}{8\pi} m^2(z)\,\Big(\varphi_{t}(z) \Phi^{[m]}_{t}(z)+
  \log |\rho_{t}(z)|^2\Big)\right]
$$
and
$$
{\tilde {\cal Z}}_t^{[m];\bar X}\equiv \exp\left[-
  \int\frac{d^2z}{8\pi} m^2(z) \left(\int_0^1 \tilde
    K_{t}^{[\sqrt{\tau} m]}(z)d\tau\right) \right].
$$

We first compute the Ito derivative of $Y_t$. From eq.(\ref{dvarphi}),
we know that $d\varphi_{t}(z)=\lambda_c \theta_{t}(z)dB_t$.  Using the
Hadamard formula, we obtain $d\Phi^{[m]}_{t}(z)= \lambda_c
\Theta^{[m]}_{t}(z)\, [dB_t -\frac{1}{2}F^{[m]}_tdt]$ with
$\lambda_c\, F^{[m]}_t= -2\int \frac{d^2z}{2\pi}
m^2(z)\,\varphi_{t}(z) \Theta^{[m]}_{t}(z)$.  The last piece of
information we need is $d\log|\rho_{t}(z)| = -\theta^2_{t}(z) dt.$ The
result for the Ito derivative of $Y_t$ is
$$ Y_t^{-1}dY_t= \half F^{[m]}_t\, dB_t - 2 N_t\, dt.$$
The drift term $-2N_tdt$ comes form the second order (crossed) term
when computing the Ito derivative of $Y_t$ and reads
$$
N_t = \int \frac{d^2z}{16\pi} m^2(z)\,\left[\lambda_c^2\theta_{t}(z)
  \Theta^{[m]}_{t}(z)-2\theta_{t}(z)^2\right].
$$
Actually the integral defining $N_t$ does not diverge at $t=0$ for
$\lambda_c^2=2$, which coincides with the value previously determined
by other considerations. Hence setting $\lambda_c =\sqrt{2}$ we have
$$
N_t=-2\int\frac{d^2z}{8\pi}\frac{d^2z'}{8\pi}
m^2(z)m^2(z')\,\theta_{t}(z) \theta_{t}(z') G^{[m]}_{t}(z',z). $$

Although it is the main result of this Section, the computation of the
derivative of $ {\tilde {\cal Z}_t^{[m];\bar X}}$ is not particularly
illuminating and we report it in appendix \ref{appendix2}.  Its Ito
derivative does not contain any $"dB_t"$ terms and there is only a
drift term. The result is:
\begin{equation}\label{dItoZ}
  d \log {\tilde {\cal Z}_t^{[m];\bar X}} = 2N_t dt,
\end{equation}
This drift compensates that of $Y_t$ and we thus find that ${\cal
  Z}_t^{[m]}$ is a local martingale:
$$ {\cal Z}_t^{[m]\,-1}\, d{\cal Z}_t^{[m]}= \half F^{[m]}_t dB_t,\quad
{\rm with}\quad F^{[m]}_t=-2\sqrt{2}\int \frac{d^2z}{4\pi} m^2(z)\,
\theta_{t}(z) \Phi^{[m]}_{t}(z)$$

In summary, ${\cal Z}_t^{[m]}$ is a local chordal SLE martingale and,
if used as a massive perturbation, the associated massive drift is
$F^{[m]}_t$, defined above. Let us note here that this drift is always
non-positive if the function $m$ is nonnegative. Indeed,
$\theta_{t}(z)=- \im \frac{2}{h_t(z)}$ is positive everywhere. As for
$m^2(z) \Phi^{[m]}_{t}(z)$, it is nonnegative on the boundary, so if
it assumed some negative values, it would have some negative absolute
minimum inside the domain (if $m$ has compact support, in particular
if the domain is bounded). At such a minimum, $-\Delta
\Phi^{[m]}_{t}(z)$ is non-positive and $m^2(z) \Phi^{[m]}_{t}(z)$ is
negative, contradicting the defining equation $(-\Delta+m^2(z))
\Phi^{[m]}_{t}(z)=0$. Hence $F^{[m]}_t$ is the integral of a
non-positive function. In particular, the driving process $\xi_t$ in
the Loewner equation is a super-martingale and we have obtained its (so
called Doob-Meyer) decomposition as a sum of a martingale and a
decreasing process explicitly. A concrete interpretation of this
decreasing process, even at small $m^2$ (the first order in
perturbation theory), would probably be of interest.

\section{Massive dipolar LERW}\label{mlerw}

In \cite{BBK} the massive drift for dipolar LERWs has been computed to
first order in the mass perturbation.  This has been done in two
different ways. The first one was by looking at the sub-interval
hitting probability i.e. the probability that a LERW from $x_0$ to the
interval $[a, b]$ ends on the sub-interval $[x,y]$. Requiring this
probability to be a martingale for massive SLE(2) gives
(perturbatively in the mass) the drift. The second approach goes
through Girsanov's formula, as explained in Section \ref{basics} in
the case of the Gaussian free field.

\subsection{Discrete massive LERW}\label{sec:dmlerw}

For the convenience of the reader, we recall here the basic
definitions for (massive) loop erased random walks.

Let us first recall the definition of a LERW.  Let us start with a
lattice of mesh $a$ embedded in a domain.  Given a path
$W=(W_0,W_1,\cdots,W_n)$ on the lattice its loop erasure $\gamma$ is
defined as follows: let $n_0=\max \{ m : W_m = W_0 \}$ and set
$\gamma_0=W_{n_0}=W_0$, next let $n_{1}=\max \{ m : W_m = W_{n_0+1}
\}$ and set $\gamma_1=W_{n_1}$, and then inductively let $n_{j+1}=\max
\{ m : W_m = W_{n_j+1} \}$ and set $\gamma_j=W_{n_j}$. This produces a
simple path $\gamma = \sL(W) = (\gamma_0, \gamma_1, \ldots, \gamma_l)$
from $\gamma_0=W_0$ to $\gamma_l=W_n$, called the loop-erasure of $W$,
but its number of steps $l$ is in general much smaller than that of
the original path $W$.  We emphasize that the starting and end points
are not changed by the loop-erasing.

We point out that the above definition of loop erasure is equivalent
to the result of a recursive procedure of chronological loop erasing:
the loop erasure of a $0$ step path $(W_0)$ is itself, $\gamma =
(W_0)$ and if the erasure of $W = (W_0,\ldots,W_m)$ is the simple path
$\sL(W) = (\gamma_0, \ldots, \gamma_l)$ then for the loop erasure of
$W' = (W_0, \ldots, W_m, W_{m+1})$ there are two cases depending on
whether a loop is formed on step $m+1$.  If $W_{m+1} \notin \{
\gamma_0 , \ldots, \gamma_l\}$ then the loop erasure of $W'$ is
$\gamma' = (\gamma_0, \ldots, \gamma_l, W_{m+1})$. But if a loop is
formed, $W_{m+1} = \gamma_k$ for some $k \leq l$ (unique because
$\gamma$ is simple), then the loop erasure of $W'$ is $\gamma' =
(\gamma_0, \ldots, \gamma_k)$.

In this paper we shall be interested in paths starting at a boundary
point ${x_0}$ and ending on a subset $S$ of the boundary of $\bD$.

Statistics of LERW is defined by associating to any simple path
$\gamma$ a weight $w_{\gamma} = \sum_{W : \sL(W) = \gamma} \mu^{|W|}$,
where the sum is over all nearest neighbor paths $W$ whose erasures
produce $\gamma$, and $|W|$ denotes the number of steps of $W$.  There
is a critical value $\mu_c$ of the fugacity at which the underlying
paths $W$ become just ordinary random walks.  The partition function
$\sum_\gamma w_\gamma$ of LERWs from $z$ to $S$ in $\bD$ can be
rewritten as a sum over walks in the domain $\bD$, started from $z$
and counting only those that exit the domain through $S$: \debut \non
Z^{\bD ; z ; S}_{\RW}
\; = \; \sum_{\substack{\gamma \textrm{ simple path} \\
    \textrm{from $z$ to $S$ in $\bD$}}} w_\gamma \; = \;
\sum_{\substack{W \textrm{ walk from}\\ \textrm{$z$ to $S$ in $\bD$}}}
\mu^{|W|} \textrm{ .}  \fin Written in terms of critical random walks,
the partition function thus reads $\EX^{z}_{\RW} \big[
(\mu/\mu_c)^{|W|} \; \ind_{W_{\tau^\RW_\bD} \in S} \big]$, where
$\tau^{\RW}_\bD$ denotes the exit time of the random walk $W$ from
$\bD$.

Critical LERW corresponds to the critical fugacity and is described by
SLE$_2$, see \cite{Sch0, LSW0, Zhan}.  For $\mu<\mu_c$ --- which is
the case we shall consider --- paths of small lengths are more
favorable and renormalization group arguments tell that at large
distances the path of smallest length dominates.  The off-critical
theory in the scaling regime corresponds to non critical fugacity
$\mu$ but approaching the critical one as the mesh size tends to zero.
At fixed typical macroscopic size, the number of steps of typical
critical random walks (not of their loop erasures) scales as $a^{-2}$,
so that the scaling limit is such that $\varrho:=-a^{-2}
\log(\mu/\mu_c)$ is finite as $a\to0$, ie.  $(\mu-\mu_c)/\mu_c\simeq
-\varrho\, a^2$ and $\varrho$ has scaling dimension $2$ and fixes a
mass scale $m^2 \simeq \varrho$ and a correlation length
$\corrl\simeq 1/m$. In this scaling limit the weights become
$(\mu/\mu_c)^{|W|}\simeq e^{-\varrho a^2|W|}$ and the random walks
converge to two dimensional Brownian motions $\Bplan$ with $a^2 |W| =
a^2 \tau^\RW_\bD$ converging to the times $\tau_\bD$ spent in $\bD$ by
$\Bplan$ before exiting.  The off-critical partition function can thus
be written as a Brownian expectation value $Z^{\bD; z; S}_{\varrho}
\longrightarrow \EX^{z}_{\BM} \big[\, e^{-\varrho \tau_\bD} \,
\ind_{\Bplan_{\tau_\bD} \in S} \, \big]$ as $a \downarrow 0$.  We may
generalize this by letting $\varrho$ vary in space: steps out of site
$w \in \bD$ are given weight factor $\mu(w) = \mu_c \; e^{-a^2
\varrho(w)}$ , in which case the partition function is an random walk
expectation value \debut \non Z^{\bD; z; S}_{\varrho} = \EX^{z}_{\RW}
\big[\, e^{-\sum_{0 \leq j < \tau^{\RW}_\bD} a^2 \varrho(W_j)} \,
\ind_{W_{\tau^{\RW}_\bD} \in S} \, \big] \underset{a \downarrow
0}{\longrightarrow} \EX^{z}_{\BM} \big[\, e^{-\int_0^{\tau_\bD}
\varrho({\Bplan}_s) \, \ud s} \, \ind_{{\Bplan}_{\tau_\bD} \in S} \,
\big] \textrm{ .}  \fin The explicit weighting by $e^{-\varrho
\tau^\RW_\bD}$ is transparent for the random walk, but becomes less
concrete for the LERW since the same path $\gamma$ can be produced by
random walks of different lengths and by walks that visit different
points.

\subsection{Continuous massive LERW}

As argued in \cite{BBK}, the field theory corresponding to (massive)
LERW is that of free massive symplectic fermions $\chi^+, \chi^-$,
with action
$$
S_{\rm sf}[\chi^\pm] = \int d^2z \Big(4\, \partial \chi^+ \bar
\partial \chi^- + m^2(z)\chi^+ \chi^- \Big)
$$
Both in the massless and in the massive case, the partition function
corresponding to dipolar SLEs can be expressed in terms of correlation
functions of boundary fields creating/annihilating the curve:
$\psi^{\pm}(x)\equiv\lim_{\delta \rightarrow 0}
\delta^{-1}\chi^\pm(x+i\delta)$.  As a consequence, the Girsanov's
martingale for massive dipolar SLE from $0$ to $[a,b]$ reads:
$$
{\cal Z}_t^{[m]}= \Big[{ \frac{{\rm Det}[-\Delta+m^2(z)]_{\bH_t}}{{\rm
      Det}[-\Delta]_{\bH_t}} }\Big]\ \frac{\langle \psi^+(\gamma_t)
  \int_{a}^{b} dx \psi^-(x) \rangle_{\bH_t}^{[m]}}{\langle
  \psi^+(\gamma_t) \int_{a}^{b} dx \psi^-(x) \rangle_{\bH_t}^{[m=0]} }
$$
where the correlation function in the numerator is computed in the
massive theory, while the one in the denominator is computed in the
massless theory. The determinants are $\zeta$-regularization of
determinants for the (massive) Laplacian with Dirichlet boundary
conditions.

By definition of the curve-creating fields $\psi^\pm$, this ratio of
correlation functions is defined by a limiting procedure:
$$
\frac{\langle\psi^+(\gamma_t) \int_{a}^{b} dx
  \psi^-(x)\rangle_{\bH_t}^{[m]}}{ \langle \psi^+(\gamma_t)
  \int_{a}^{b} dx \psi^-(x) \rangle_{\bH_t}^{[m=0]} }= \lim_{z\to
  \gamma_t} \frac{ \Psi^{[m]}_{t,[a,b]}(z) }{ \Psi^{[0]}_{t,[a,b]}(z)}
$$
where $\Psi^{[m]}_{t,[a,b]}(z)= \langle \chi^+(z) \int_{a}^{b} dx
\psi^-(x) \rangle_{\bH_t}^{[m]}$.  By construction, $\Psi^{[m]}_{t,[a,
  b]}(z)$ satisfies the massive Laplace equation $(-\Delta+m^2(z))
\Psi^{[m]}_{t, [a, b]}(z) =0$ with specific boundary conditions.  This
allows us to write it in terms of the massless correlation function
$\Psi^{[0]}_{t,[a,b]}(z)$ and of the massive Green function
$G_t^{[m]}(z,w)$.  We may then take the limit $z\to\gamma_t$ as the
limit $g_t(z)\to \xi_t$ so that this ratio becomes
$$
\frac{\langle\psi^+(\gamma_t) \int_{a}^{b} dx
  \psi^-(x)\rangle_{\bH_t}^{[m]}}{ \langle \psi^+(\gamma_t)
  \int_{a}^{b} dx \psi^-(x) \rangle_{\bH_t}^{[m=0]} }=
\frac{\Gamma^{[m]}_{t,[a,b]} }{ \Gamma^{[0]}_{t,[a,b]} }
 $$
 with
 \begin{eqnarray}
   \Gamma^{[m]}_{t,[a,b]}=\Gamma^{[0]}_{t,[a,b]}-\int\frac{d^2z}{4\pi}
   m^2(z)\,\Theta^{[m]}_t(z)\, \Psi^{[0]}_{t,[a,b]}(z)
   \label{gamma_tm}
 \end{eqnarray}
 where $\Theta^{[m]}_t(\cdot)$ is the massive Poisson kernel.  From
 this expression, we see that $\Gamma^{[m]}_{t,[a,b]}$ depends
 explicitly on $\xi_t$, on $a_t=g_t(a)$ and $b_t=g_t(b)$ and on $t$.
 When computing its Ito derivative, only the explicit dependence on
 $\xi_t$ contributes to the "$dB_t$" term, the rest contributes to the
 "$dt$" term. See Appendix \ref{ito_gamma} for the definition of
 $\Gamma^{[0]}_{t,[a,b]}$ and $\Psi^{[0]}_{t,[a,b]}(z)$ and more
 details.

 From Girsanov's theorem, $\sqrt{2}{\cal Z}_t^{[m]\, -1} d {\cal
   Z}_t^{[m]}$ gives the additional drift due to the massive
 perturbation. As explained in Section 2, the critical drift
 $F^0_{t,[a,b]}$ derive from the critical chordal SLE martingale
 $\Gamma^{[0]}_{t,[a,b]}$, which intertwines dipolar and chordal SLEs.
 Therefore, ${\cal Z}_t^{[m]}$ is a dipolar martingale whenever
 $\tilde {\cal Z}_t^{[m]}\equiv {\cal Z}_t^{[m]}\
 \Gamma^{[0]}_{t,[a,b]}$ is a chordal martingale.  Explicitly:
 \begin{eqnarray}
   \tilde {\cal Z}_t^{[m]}= \Big[{
     \frac{{\rm Det}[-\Delta+m^2(z)]_{\bH_t}}{{\rm Det}[-\Delta]_{\bH_t}}
   }\Big] \,\Gamma^{[m]}_{t,[a,b]}
   \label{Ztilde}
 \end{eqnarray}
 Let $\tilde B_t$ be the Brownian motion associated to the critical
 chordal LERW (not that of dipolar LERW).  The massive dipolar drift
 is then:
$$d\xi_t = \sqrt{2} dB^{[m]}_t +  F^{[m]}_{t,[a,b]}\, dt, \quad
\sqrt{2}\,\tilde {\cal Z}_t^{[m]\,-1}d\tilde {\cal Z}_t^{[m]} =
F^{[m]}_{t,[a,b]} d\tilde B_t,
$$ 
where $B^{[m]}_t$ is a Brownian motion with respect to the
off-critical measure $\EX^{[m]}[\cdots]$.

In order to avoid infinities appearing similarly as for the Gaussian
free field, we consider the Ito derivative of the product
$\Gamma^{[m]}_{t,[a,b]}\, e^{J_t}$ where
$$
J_t=\int \frac{d^2z}{4\pi} m^2(z) \log|\rho_t(z)|^2.
$$
The computation of this derivative which is again based on the
Hadamard formula is reported in Appendix \ref{ito_gamma}. It reads:
\begin{equation}\label{dgamma}
  d\left[\Gamma^{[m]}_{t,[a,b]}e^{J_t} \right] =
  \Gamma^{[m]}_{t,[a,b]}\,e^{J_t}\left[\sqrt{2}\, \Big(\partial_{\xi_t} 
    \log\Gamma^{[m]}_{t,[a,b]}\Big)\, d\tilde B_t + 4N_tdt\right]
\end{equation}
where
$$
N_t= \int \frac{d^2z}{8\pi} m^2(z)
[\Theta_t^{[m]}(z)\theta_t(z)-\theta^2_t(z)].
$$ 
is the same quantity that we have encountered in Section
\ref{sle4mart}. The key point here is that the drift term in
$d[\Gamma^{[m]}_{t,[a,b]}e^{J_t}]$ is $4N_t\,
\Gamma^{[m]}_{t,[a,b]}e^{J_t}$.  The derivative of the ratio of
functional determinants has already been computed in Section
\ref{sle4mart} with result:
$$
d\left[\log \left[e^{-J_t}\, \frac{{\rm
        Det}(-\Delta+m^2(z))_{\bH_t}}{{\rm Det}(-\Delta)_{\bH_t}}
  \right] \right] = - 4N_t dt.
$$
This drift cancels exactly the one coming from
$d\left[\Gamma^{[m]}_{t,[a,b]}\,e^{J_t} \right]$. In conclusion we
find:
\begin{equation*} {\cal Z}_t^{[m]\,-1} d {\cal Z}_t^{[m]} = \sqrt{2}\,
  \Big( \partial_{\xi_t} \log\Gamma^{[m]}_{t,[a,b]}\Big)d\tilde B_t,
\end{equation*}
which means that ${\cal Z}_t^{[m]}$ is a (local) martingale for the
critical chordal measure and the off-critical drift reads
\begin{equation}
  F^{[m]}_{t,[a,b]}=2\,\Big(\partial_{\xi_t}\log\Gamma^{[m]}_{t,[a,b]}\Big).
  \label{drift_lerw}
\end{equation}

\subsection{Massive symplectic correlation functions}
We now show that, as expected from basic rules of statistical
mechanics, ratio of correlation functions of massive symplectic
fermions
$$ \frac{\vev{\psi^+(\gamma_t){\cal O}}^{[m]}_{\bH_t}}{
  \vev{\psi^+(\gamma_t)\int_a^b dx\psi^-(x)}^{[m]}_{\bH_t}}$$ are
local martingales for massive dipolar SLE(2).  These ratios are
defined by a limiting procedure which can be written as:
$$
\lim_{z\to\gamma_t}\frac{\vev{\chi^+(z){\cal O}}^{[m]}_{\bH_t}}{
  \vev{\chi^+(z)\int_a^bdx\psi^-(x)}^{[m]}_{\bH_t}}$$ As in the
previous Section, this limit is taken by letting $g_t(z)$ approach
$\xi_t$, which leads us to write:
$$\frac{\vev{\psi^+(\gamma_t){\cal O}}^{[m]}_{\bH_t}}{
  \vev{\psi^+(\gamma_t)\int_a^b dx\psi^-(x)}^{[m]}_{\bH_t}}=
\frac{\vev{\sO}^{[m]}_t}{\Gamma^{[m]}_{t,[a,b]}}$$ This serves as
definition for $\vev{\sO}^{[m]}_t$.

To prove that these ratio are local martingales, we have to compute
their Ito derivatives with the massive drift.  These can be presented
in the following form:
\begin{eqnarray*}
  d\left[\langle\sO\rangle^{[m]}_t e^{J_t}\right] &=& \langle\sO
  \rangle^{[m]}_te^{J_t}\left[X_t^\sO  d \tilde B_t + R_t^\sO dt\right]\\
  &=& \langle \sO\rangle^{[m]}_t e^{J_t}\left[X_t^\sO
    (d B_t^{[m]}+  \frac{1}{\sqrt{2}} F^{[m]}_{t,[a,b]} dt ) + R_t^\sO
    dt \right]. 
\end{eqnarray*}
Combining this equation with the formula (\ref{dgamma}) of the Ito
derivative of $\Gamma^{[m]}_{t,[a,b]} e^{J_t}$and $d\tilde
B_t=dB^{[m]}_t+\frac{1}{\sqrt{2}} F^{[m]}_{t,[a,b]} dt$, we obtain the
Ito derivative of the ratio $\langle \sO \rangle^{[m]}_t
/\Gamma^{[m]}_{t,[a,b]}$:
$$
d\left[\langle \sO \rangle^{[m]}_t /\Gamma^{[m]}_{t,[a,b]}\right] =
[\langle \sO\rangle^{[m]}_t /\Gamma^{[m]}_{t,[a,b]}]\ [(X_t^\sO -
\frac{1}{\sqrt{2}} F^{[m]}_{t,[a,b]}) dB_t^{[m]} +(R_t^\sO - 4N_t)dt]
$$
The condition for $\langle \sO \rangle^{[m]}_t/\Gamma^{[m]}_{t,[a,b]}$
to be a martingale for massive dipolar SLE(2) is thus
\begin{equation}\label{requir}
  R_t^\sO = 4N_t ~~~~\textrm{independently of}~~\sO. 
\end{equation}
Let us check it in few examples.
\subsubsection*{Examples}
\begin{itemize}
\item Consider $\Gamma^{[m]}_{t,[x,y]}$ for two points $x,y$ different
  from $a,b$. From eq.(\ref{dgamma}) we know that:
$$
d\left[\Gamma^{[m]}_{t,[x,y]}e^{J_t} \right]= \Gamma^{[m]}_{t,[x,y]}
e^{J_t}\left[\sqrt{2} \left(\partial_{\xi_t}
    \log\Gamma^{[m]}_{t,[x,y]}\right) d\tilde B_t + 4N_tdt]\right]
$$
Therefore $\Gamma^{[m]}_{t,[x,y]}/\Gamma^{[m]}_{t,[a,b]}$ is a
$P_m$-martingale. Actually such a martingale has a simple
interpretation when the points $x$ and $y$ belong to the interval
$[a,b]$. In such a case the ratio
$\Gamma^{[m]}_{0,[x,y]}/\Gamma^{[m]}_{0,[a,b]}$ gives the probability
that a massive LERW started in the origin and conditioned to end on
the interval $[a,b]$ hits the sub-interval $[x,y]$, see
ref.\cite{BBK}.

\item Consider $\sO =\chi_-(z) $, then $\vev{\sO}^{[m]}_t=
  \Theta_t^{[m]}(z)$ is the Poisson kernel. In Appendix
  \ref{ito_gamma} we compute its Ito derivative and the result is
$$
d\left[ \Theta_t^{[m]}(z) e^{J_t}\right] = e^{J_t} \left[Q_t^{[m]}(z)
  \sqrt{2} d\tilde B_t + 4\Theta_t^{[m]}(z) N_t\right] dt.
$$
Thus $\Theta_t^{[m]}(z)/\Gamma^{[m]}_{t,[a,b]}$ is a $P_m$ SLE(2)
martingale.

\item We generalize the previous two examples by considering an
  arbitrary product of fermions
$$\sO =
\prod_{j=1}^{N+1} \chi_-(z_j) \prod_{k=1}^{N}\chi_+(w_k).$$ The total
charge has to be $-1$ as $\psi^+$ carries charge $+1$.  Using Wick's
theorem we have
$$
\langle \sO \rangle^{[m]}_t = \det \left[
  \begin{array}{cccc}
    G_t^{[m]}(z_1,w_1) &\dots &G_t^{[m]}(z_N,w_1) & \Theta_t^{[m]}(w_1)\\
    G_t^{[m]}(z_1,w_2) &\dots &G_t^{[m]}(z_N,w_2) & \Theta_t^{[m]}(w_2)\\
    \vdots  & \ddots & \vdots & \vdots\\
    G_t^{[m]}(z_1,w_{N+1}) &\dots &G_t^{[m]}(z_N,w_{N+1}) &
    \Theta_t^{[m]}(w_{N+1}) 
  \end{array}
\right]
$$
Looking at the drift term of the Ito derivative of $ \langle \sO
\rangle^{[m]}_t e^{J_t}$ we notice that there are no contributions
coming from the second order term, since $d G_t^{[m]}(z,w)$ has no
term proportional to $d\tilde B_t$.  The first order terms are of two
kinds. The first one, which is of the expected form $4\langle \sO
\rangle^{[m]}_t e^{J_t} N_t $, comes from the derivative of the last
column.  There are other contributions coming from the derivative of
each other column. Thanks to the Hadamard formula, the contribution of
the derivative of the $j$-th column is proportional to:
$$
\det\left[
  \begin{array}{ccccc}
    G_t^{[m]}(z_1,w_1) &\dots & \Theta_t^{[m]}(z_j)\Theta_t^{[m]}(w_1)
    &\dots & \Theta_t^{[m]}(w_1)\\ 
    G_t^{[m]}(z_1,w_2) &\dots & \Theta_t^{[m]}(z_j) \Theta_t^{[m]}(w_2) &
    \dots & \Theta_t^{[m]}(w_2)\\ 
    \vdots  & \ddots & \vdots & \ddots & \vdots\\
    G_t^{[m]}(z_1,w_{N+1}) &\dots & \Theta_t^{[m]}(z_j)\Theta_t^{[m]}(w_{N+1})
    &\dots  & \Theta_t^{[m]}(w_{N+1}) 
  \end{array}
\right].
$$
This however is zero because the last and the $j$-th columns are
proportional.
\end{itemize}

We therefore conclude that $\langle \sO \rangle^{[m]}_t$ satisfy
conditions (\ref{requir}) and thus that all correlation functions
$\langle \sO \rangle^{[m]}_t/\Gamma^{[m]}_{t,[a,b]}$ are $P_m$ (local)
martingales. This is analogue to a perfect matching but between
(massive) symplectic fermions and (massive) LERW.

\appendix

\section{Computation of determinant ratio}\label{appendix1}
In this appendix we compute the ratio of spectral determinants that we
use in Sections \ref{sle4mart} and \ref{mlerw}:
$$
\frac{{\rm Det}[-\Delta+m^2(z)]_{\bH_t}}{{\rm Det}[-\Delta]_{\bH_t}}
$$
We define the determinant of a self-adjoint elliptic operator $\cal D$
defined on a domain $\cal M$ through the $\zeta$-function
regularization.  Let $\zeta_{\cal D}(s)$ be defined as
\begin{equation}\label{def-det}
  \zeta_{\cal D}(s) = \frac{1}{\Gamma(s)}\int_0^\infty t^{s-1} {\rm
    Tr}(e^{{-\cal D} t})dt
\end{equation}  
where $e^{{-\cal D} t}$ is the heat kernel associated to the operator
$\cal D$. The integral defining the $\zeta$-function is convergent
only for $\re(s)> s_0> 0$ but the $\zeta$-function itself can be
analytically continued in $s=0$ where it is holomorphic.  Then the
prescription for the determinant is
\begin{equation}
  \log{\rm Det}[{\cal D}] \equiv - \zeta'_{\cal D}(0).
\end{equation} 
In our case we are interested in getting a difference of logarithms of
determinants
$$
\log\left[\frac{{\rm Det}[-\Delta +m^2]_{\bH_t}}{{\rm
      Det}[-\Delta]_{\bH_t}} \right] = -
\zeta'_{-\Delta+m^2}(0)+\zeta'_{-\Delta}(0) =-\int_0^1
d\tau\frac{d}{d\tau} \zeta'_{-\Delta+\tau m^2}(0).
$$
We are going to evaluate $\frac{d}{d\tau}\zeta_{-\Delta+\tau m^2}(s)$
for $s$ close to zero. Taking the derivative is easy because
$\frac{d}{d\tau}{\rm Tr}\Big(e^{(\Delta-\tau m^2)t}\Big) =-{\rm
  Tr}\Big(m^2\,e^{(\Delta-\tau m^2)t}\Big)$.  In order to perform the
analytic continuation which gives the $\zeta$-function in $0$ we
separate the integral in eq.(\ref{def-det}) in two parts introducing a
cut-off $\epsilon$:
\begin{equation*}
  \frac{d}{d\tau}\zeta_{-\Delta +\tau m^2}(s) 
  = -\frac{1}{\Gamma(s)}\left(\int_0^\epsilon dt
    +\int_\epsilon^\infty dt \right)t^{s}{\rm
    Tr}\Big(m^2\,e^{(\Delta-\tau m^2)t}\Big)  
\end{equation*}
This equation is true for any $\epsilon$ but we shall take the limit
$\epsilon\to0$ after having implemented the analytic continuation.
The second integral can be directly continued to $s$ around $0$ since
the divergence has been cut off.  The first integral of course cannot
be computed for $s$ around $0$ but, since we are going to send
$\epsilon \rightarrow 0$, we can compute it using the small time
expansion of the heat kernel \cite{oper_ellip}.  So let
$P_t^{[\sqrt{\tau}m]}\equiv e^{(\Delta-\tau m^2)t}$. For small $t$ we
have the expansion:
$$P_t^{[\sqrt{\tau}m]}(z,w)=
P_t^{[0]}(z,w)\Big(1+\sum_{j\geq1}t^{j/2}\phi_j(z,w)\Big)$$ with
$P_t^{[0]}$ the massless heat kernel with Dirichlet boundary
conditions.  Inserting this expansion in the first integral and using
the fact that along the diagonal $P_t^{[0]}(z,z)=\frac{1}{4\pi t}$, up
to exponentially small term as $t\to 0$, gives:
$$\int_0^\epsilon dt\, t^{s-1}{\rm Tr}\big(m^2\,e^{(\Delta -\tau m^2)t}\Big) 
=\frac{\epsilon^s}{s}\int\frac{d^2z}{4\pi}m^2(z)+ \cdots$$ where the
dots refer to sub-leading terms in $\epsilon$.  Taking the derivative
of the $\zeta$-function w.r.t. $s$ (recall that
$s\Gamma(s)=\Gamma(s+1)$) we arrive at
$$
\frac{d}{d\tau} \zeta'_{-\Delta+\tau m^2}(0) =\lim_{\epsilon
  \rightarrow 0} \left[(\Gamma'(1)-\log\epsilon)\int\frac{d^2 z}{4\pi}
  m^2(z) - \int_\epsilon^\infty {\rm Tr}(m^2 e^{(\Delta-\tau m^2)t})dt
\right]
$$
\begin{eqnarray}\label{derivative-zeta}
  =\lim_{\epsilon \rightarrow 0}
  \left[(\Gamma'(1)-\log\epsilon)\int\frac{d^2 z}{4\pi} m^2(z) 
    -   {\rm Tr}\left(\frac{1}{-\Delta+\tau m^2}m^2 
      e^{(\Delta-\tau m^2)\epsilon}\right)   \right]
\end{eqnarray}
It is now again a matter of small time expansion of the heat kernel.
We have:
$$
{\rm Tr}\left(\frac{1}{-\Delta+\tau m^2}m^2 e^{(\Delta -\tau
    m^2)\epsilon}\right) = \int \frac{d^2zd^2z'}{4\pi}
G^{[\sqrt{\tau}m]}_{t}(z',z) m^2(z)P^{[\sqrt{\tau}m]}_\epsilon(z,z')
$$
We compute this integral by adding and subtracting $\log|z-z'|^2$ to
$G^{[\sqrt{\tau}m]}_{t}(z',z)$ and splitting the integral into two
integrals.  The first one involves
$G^{[\sqrt{\tau}m]}_{t}(z',z)+\log|z-z'|^2$. There we can directly
take the limit $\epsilon\to0$. Using the fact that
$\lim_{\epsilon\to0}P^{[\sqrt{\tau}m]}_\epsilon(z,z')=\delta(z,z')$ we
get
$$\int\frac{d^2 z}{4\pi} m^2(z)\lim_{z'\to z}
\Big(G^{[\sqrt{\tau}m]}_{t}(z',z)+\log|z-z'|^2\Big)$$ By definition
the last term is $\langle \bar
X^2(z)\rangle_{\bH_t}^{[\sqrt{\tau}m]}$.  The second integral involves
$\log|z-z'|^2$. In the limit $\epsilon\to0$ of that integral we can
replace $P^{[\sqrt{\tau}m]}_\epsilon(z,z')$ by
$P^{[0]}_\epsilon(z,z')$. The integral over $z'$ can then be exactly
evaluated to give
$$ \int d^2z' \log|z-z'|^2\,P^{[0]}_\epsilon(z,z')
= (\log(4\epsilon) +\Gamma'(1))\int\frac{d^2 z}{4\pi} m^2(z)+\cdots$$
Putting everything together we get for ${\rm
  Tr}\left(\frac{1}{-\Delta+\tau m^2}m^2 e^{(\Delta-\tau
    m^2)\epsilon}\right)$
$$
=-(\log(4\epsilon) +\Gamma'(1))\int\frac{d^2 z}{4\pi} m^2(z) +\int
\frac{d^2z}{4\pi} m^2(z)\, \langle \bar
X^2(z)\rangle_{\bH_t}^{[\sqrt{\tau}m]} + O(\epsilon),
$$
where $\langle \bar X^2(z)\rangle_{\bH_t}^{[\sqrt{\tau}m]}$ is given
exactly by the point splitting regularization.  Once we substitute
this expression for the trace into eq.(\ref{derivative-zeta}), we get:
\begin{eqnarray}\label{zetadet_fin}
  \frac{d}{d\tau}\log\left[\frac{{\rm Det}[-\Delta +\tau
      m^2]_{\bH_t}}{{\rm Det}[-\Delta]_{\bH_t}} \right] 
  &=&-\frac{d}{d\tau} \zeta'_{-\Delta +\tau m^2}(0) \\
  && \hskip -2.0 truecm = \int \frac{d^2z}{4\pi} m^2(z)\, 
  \langle \bar X^2(z)\rangle_{\bH_t}^{[\sqrt{\tau}m]}
  + {\rm const.} \int \frac{d^2z}{4\pi} m^2(z) \nonumber
\end{eqnarray}
Up to the irrelevant term proportional to $\int \frac{d^2z}{4\pi}
m^2(z)$ that we can and shall ignore, this coincides with the naive
field theory derivation, eq.(\ref{dlogZtau}).

\section{Derivative of $\log {\tilde {\cal Z}_t^{[m];\bar X}}$}\label{appendix2}
Here we compute the derivative of $\log {\tilde {\cal Z}_t^{[m];\bar
    X}}$:
$$
d\log {\tilde {\cal Z}_t^{[m];\bar X}} = -\half \int_0^1 d\tau
\left(\int_{\bH_t}\frac{d^2z}{4\pi}m^2(z)\, d K_{t}^{[\sqrt{\tau}
    m]}(z)\right).
$$
We are going to show that $dK_{t}^{[\sqrt{\tau} m]}(z)$ can be written
as a total derivative w.r.t. $\tau$:
\begin{equation}\label{tot-der}
  d K_{t}^{[\sqrt{\tau} m]}(z) = 2\left(\int \frac{d^2z'}{4\pi}
    \frac{d}{d\tau}\left[ \tau^2 m^2(z')\,\theta_{t}(z) 
      \theta_{t}(z') G^{[\sqrt{\tau}m]}_{t}(z',z)\right]   \right) dt.
\end{equation}
Indeed, on one hand we can use the expression for $K_{t}^{[\sqrt{\tau}
  m]}(z)$ given in eq.(\ref{Ktilde}) and the massless and massive
Hadamard formulas to write the left hand side of eq.(\ref{tot-der}) as
\begin{eqnarray*}
  d K_{t}^{[\sqrt{\tau} m]}(z)=\,
  2\int_{\bH_t}\frac{d^2z'}{4\pi}
  \,\Theta^{[\sqrt{\tau}m]}_{t}(z)
  \Theta^{[\sqrt{\tau}m]}_{t}(z')\tau m^2(z') G^{[0]}_{t}(z',z)dt\\ 
  + 2\int_{\bH_t}\frac{d^2z'}{4\pi}
  \,G^{[\sqrt{\tau}m]}_{t}(z,z')\tau m^2(z')
  \theta_{t}(z')\theta_{t}(z)dt.   
\end{eqnarray*}
On the other hand, if we develop the derivative w.r.t.  $\tau$ on the
right hand side of eq.(\ref{tot-der}) we get
$$4\int \frac{d^2z'}{4\pi}\tau 
m^2(z')\,\theta_{t}(z)\theta_{t}(z') G^{[\sqrt{\tau}m]}_{t}(z',z)\,
dt$$
$$-2\int \frac{d^2z'}{4\pi}\frac{d^2z''}{4\pi}\tau^2 
m^2(z')\,\theta_{t}(z) \theta_{t}(z')
G^{[\sqrt{\tau}m]}_{t}(z',z'')m^2(z'')G^{[\sqrt{\tau}m]}_{t}(z'',z)\,$$
$$=2\int \frac{d^2z'}{4\pi}\tau m^2(z')\,\theta_{t}(z) 
\theta_{t}(z') G^{[\sqrt{\tau}m]}_{t}(z',z)\, dt $$
$$+2\int \frac{d^2z'}{4\pi}\tau 
m^2(z')\,\Theta^{[\sqrt{\tau}m]}_{t}(z) \theta_{t}(z')
G^{[\sqrt{\tau}m]}_{t}(z',z)\, dt.$$ Eq.(\ref{tot-der}) follows from
the fact that
$$\int d^2z'\Theta^{[\sqrt{\tau}m]}_{t}(z')\tau m^2(z') G^{[0]}_{t}(z',z)
= \int d^2z'\theta_{t}(z') \tau m^2(z')G^{[\sqrt{\tau}m]}_{t}(z',z)$$
Once we have this relation we can plug it into the equation for the
derivative of $d\log {\tilde {\cal Z}_t^{[m];\bar X}}$ and we get
\begin{equation}\label{dlog}
  d\log {\tilde {\cal Z}_t^{[m];\bar X}}=-4\int
  \frac{d^2z}{8\pi}\frac{d^2z'}{8\pi} m^2(z)m^2(z')\,\theta_{t}(z)
  \theta_{t}(z') G^{[m]}_{t}(z',z) dt.
\end{equation}
The right hand side is nothing else than $2N_t dt$.  This proves
eq.(\ref{dItoZ}).

\section{LERW: Ito derivatives}\label{ito_gamma}
In this appendix we present some explicit formul\ae\ which are
instrumental to the computations performed in Section \ref{mlerw}. We
comment also about some apparent divergences present in the
computation of the Ito derivative of $\Gamma^{[m]}_{t,[a,b]}$ and of
other quantities.

Recall the definition:
$$
\Psi^{[m]}_{t,[a, b]}(z)= \langle \chi^+(z) \int_{a}^{b} dx
\psi^-(x)\rangle_{\bH_t}^{[m]}.
$$
By construction and an appropriate choice of normalization,
$\Psi^{[m]}_{t,[a, b]}(z)$ satisfies the massive Laplace equation
$(-\Delta+m^2(z)) \Psi^{[m]}_{t, [a, b]}(z) =0$ with boundary
conditions: $\Psi^{[m]}_{t, [a,b]}(z)=\pi$ when $z\in [a,b]$, instead
$\Psi^{[m]}_{t, [a, b]}(z)=0$ when $z$ lies outside the interval
$[a,b]$.  We may write it in terms of the massive Green function:
$$\Psi^{[m]}_{t,[a,b]}(z)= \Psi^{[0]}_{t,[a,b]}(z)-\frac{1}{4\pi}
G^{[m]}_t(z,\cdot)\star m^2(\cdot) \Psi^{[0]}_{t,[a,b]}(\cdot)$$ with
($a_t=g_t(a)$ and $b_t=g_t(b))$
$$ \Psi^{[0]}_{t,[a,b]}(z)=\im \log\Big(\frac{g_t(z)-a_t}{g_t(z)-b_t}\Big).$$

 From the relation between $\psi^\pm$ and $\chi^\pm$, it follows that
$\Gamma^{[m]}_{t,[a,b]}$ is defined from a limiting procedure from
$\Psi^{[m]}_{t, [a,b]}(z)$. We set:
$$
\Gamma^{[m]}_{t,[a,b]} = \lim_{\delta \to 0} \frac{1}{2\delta}
\Psi^{[m]}_{t, [a,b]}(z )\Big\vert_{g_t(z)=\xi_t+i\delta}.
$$
In the massless case we have
$$
\Gamma^{[0]}_{t,[a,b]}= \frac{(a_t-b_t)}{(\xi_t-a_t)(\xi_t-b_t)}
$$
As usual we can write the massive solutions in terms of the massless
ones and of the massive propagator. This gives
\begin{eqnarray}\label{expressG}
  \Gamma^{[m]}_{t,[a,b]} &=& \Gamma^{[0]}_{t,[a,b]}
  -\frac{1}{4\pi} \Theta^{[m]}_t(\cdot)\star
  m^2(\cdot)\Psi^{[0]}_{t,[a,b]}(\cdot)  
\end{eqnarray}
with $\Theta^{[m]}_t(z)=\theta_t(z) -
\frac{1}{4\pi}G^{[m]}_t(z,\cdot)\star m^2(\cdot)\theta_t(\cdot)$.

We can now compute the Ito derivatives.  The ingredients we need are:
\begin{eqnarray*}
  d \Gamma^{[0]}_{t,[a,b]} &=& \Gamma^{[0]}_{t,[a,b]}\,
  F^0_{t,[a,b]}\sqrt{2} d\tilde B_t,\\
  d\Psi^{[0]}_{t,[a,b]}(z) &=&- 2 \theta_t(z)\, \Gamma^{[0]}_{t,[a,b]} dt,\\
  d\theta_t(z) &=& Q^{[0]}_t(z) \sqrt{2}d\tilde B_t,
\end{eqnarray*}
with $Q^{[0]}_t(z)= -2\im \frac{1}{(z_t-\xi_t)^2}$ .  The last
equation and the Hadamard formula
$dG^{[m]}_t(z,w)=-2\Theta_t^{[m]}(z)\Theta_t^{[m]}(w)dt$ imply that:
$$d\Theta_t^{[m]}(z) = Q^{[m]}_t(z) \sqrt{2}d\tilde B_t +
4\Theta_t^{[m]}(z)\, \hat N_t\,dt,$$ with
$$Q^{[m]}_t(z)= Q^{[0]}_t(z)- \frac{1}{4\pi}G_t^{[m]}(z,\cdot)\star
m^2(\cdot)Q^{[0]}_t(\cdot)$$
$$\hat N_t = \int \frac{d^2z}{8\pi} m^2(z)\Theta_t^{[m]}(z)\theta_t(z).$$
Ito differentiating eq.(\ref{expressG}) and putting all these pieces
together we find
$$
d\left[ \Gamma^{[m]}_{t,[a,b]}\right] = \Big(\Gamma^{[0]}_{t,[a,b]}
F^0_{t,[a,b]}- \frac{1}{4\pi}Q_t^{[m]}(\cdot)\star m^2(\cdot)
\Psi^{[0]}_{t,[a,b]}(\cdot) \Big) \sqrt{2} d\tilde B_t
+4\Gamma^{[m]}_{t,[a,b]} \hat N_t dt
$$
By construction
$$\partial_{\xi_t} \Gamma^{[m]}_{t,[a,b]}= \Gamma^{[0]}_{t,[a,b]}
F^0_{t,[a,b]}- \frac{1}{4\pi}Q_t^{[m]}(\cdot)\star m^2(\cdot)
\Psi^{[0]}_{t,[a,b]}(\cdot) $$ Again the key point is that the drift
term in the previous equation is $4\Gamma^{[m]}_{t,[a,b]}\, \hat N_t
dt$.

Here we encounter an unpleasant problem. Indeed $\hat N_t$ naively
diverges as $t\rightarrow 0$. In order to avoid such a problem one can
instead consider $\Gamma^{[m]}_{t,[a,b]} e^{J_t} $, where we recall
the definition of $J_t$:
$$
J_t=\int \frac{d^2z}{4\pi} m^2(z) \log|\rho_t(z)|^2.
$$
Recall that $d\log|\rho_t(z)|=-\theta_t(z)^2dt$.  Taking now the Ito
derivative of $\Gamma^{[m]}_{t,[a,b]} e^{J_t}$, we get for $t>0$
\begin{eqnarray*}
  d\left[ \Gamma^{[m]}_{t,[a,b]} e^{J_t} \right] =
  \Big(e^{J_t}\, \partial_{\xi_t}\Gamma^{[m]}_{t,[a,b]} \Big)\, \
  \sqrt{2} d\tilde B_t +4\Gamma^{[m]}_{t,[a,b]}e^{J_t}\, N_t dt
\end{eqnarray*}
with
$$
N_t=\int \frac{d^2z}{8\pi} m^2(z)
\left[\Theta_t^{[m]}(z)\theta_t(z)-\theta^2(z)\right].
$$
This quantity is now finite as $t\rightarrow 0$.  This proves
eq.(\ref{dgamma}).

\bigskip

\emph{Acknowledgments:} We wish to thank S. Smirnov and P.  Wiegmann
for discussions.

Our work is supported by ANR-06-BLAN-0058-01 (D.B. and L.C.),
ANR-06-BLAN-0058-02 (M.B.)  and ENRAGE European Network
MRTN-CT-2004-5616.

\end{document}